\documentclass[%
 preprint, 
 amsmath,amssymb,
 aps, physrev,
]{revtex4-2}

\usepackage{graphicx}
\usepackage{dcolumn}
\usepackage{bm}
\usepackage{hyperref}
\hypersetup{
    colorlinks=true,
    linkcolor=blue,
    citecolor=blue,
    urlcolor=blue
}

\begin{document}

\preprint{}

\title{\textbf{Effect of Spatially Heterogeneous Mucin Coverage on Tear Film Stability and Rupture} 
}%

\author{Deepak Kumar}
\author{Pushpavanam S}%
 \email{Contact author: spush@iitm.ac.in}
\affiliation{Department of Chemical Engineering, Indian Institute of Technology Madras, Chennai, 600036, India}%

\date{\today}

\begin{abstract}
Clinical observations of dry eyes reveal that tear film breakup is associated with spatial variations in corneal wettability arising from non-uniform mucin coverage. Motivated by these observations, we develop a thin-film model to investigate the influence of heterogeneous wettability on tear film stability. Heterogeneity in mucin coverage is incorporated through variations in the Hamaker constant and slip length along the corneal surface. Two representative forms of spatial heterogeneity are considered: a periodic step variation representing sharply localised mucin-deficient patches and a smoothly varying sinusoidal distribution representing gradual changes in glycocalyx. The steady states are obtained by a balance between capillary and van der Waals forces. A linear stability framework based on Floquet-Bloch theory and a discretised eigenvalue approach is developed to account for the periodic coefficients in the linearised equations. We show that heterogeneous wettability induces coupling between perturbation modes. The most unstable wavenumber and the maximum growth rate decrease with increasing mucin coverage fraction. However, both increase with increasing Hamaker constant contrast between mucin-rich and mucin-deficient regions. Nonlinear simulations reveal that rupture preferentially localises within mucin-deficient regions irrespective of the initial film thickness. The rupture location is governed by the spatial distribution of disjoining pressure rather than the initial perturbation. The predicted rupture dynamics are consistent with clinical observations where rupture location is invariant and the rupture times obtained from the model are in good agreement with clinically reported values. These findings demonstrate that spatial heterogeneity in wettability plays a decisive role in tear film instability and must be incorporated in tear film dynamics models.
\clearpage
\end{abstract}

\maketitle



\section{Introduction}
\label{sec:Introduction}

\begin{figure}
  \centering
  \includegraphics[width=1\textwidth]{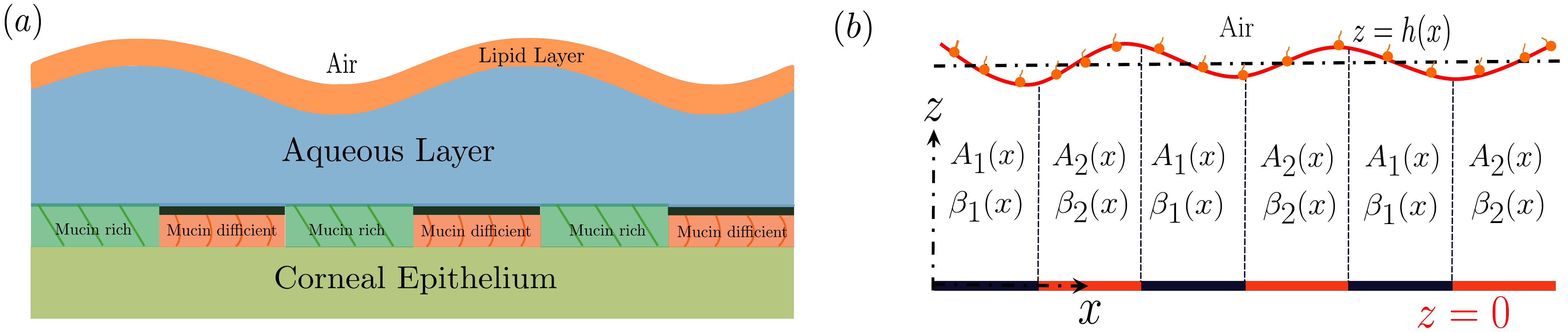}
  \caption{$(a)$ Illustration of the tear film structure over the heterogeneous mucin covering epithelial surface $(b)$ Mathematical model representation of a thin tear film over a corneal surface with spatially heterogeneous wettability.}
  \label{fig:1}
\end{figure}

 The tear film is a thin liquid layer that coats the ocular surface and plays a crucial role in protecting the cornea. It is commonly described as a tri-layered structure, as illustrated in FIG \ref{fig:1}(a). It consists of a lipid layer, aqueous layer and mucin layer. The lipid layer is secreted by the Meibomian glands. It suppresses evaporation and enhances interfacial stability by modulating surface tension \citep{bron2004functional,craig1997importance,mcculley1997compositional,zhang2003analysis}. The aqueous layer is produced primarily by the lacrimal glands. It constitutes the bulk of the tear film and supplies oxygen, nutrients, and antimicrobial agents to the cornea \citep{mcdermott2013antimicrobial}. The innermost mucin layer is produced by conjunctival goblet cells and consists of membrane-associated glycoproteins expressed on the apical surface of corneal epithelial cells \citep{davidson2004tear}. This layer enhances wettability by reducing the intrinsic hydrophobicity of the epithelial surface and promotes adhesion between the tear film and the cornea \citep{cho1991stability,hodges2013tear}. The membrane-associated mucins form a hydrated, soft, and flexible glycocalyx that facilitates partial slip of the tear film over the corneal surface \citep{braun2007model}. \\
The stability of the tear film is governed by a balance between capillary forces, intermolecular interactions and the interfacial effects associated with the lipid layer. Disruption of this balance leads to tear film breakup which is a central pathophysiological feature of dry eye disease (DED). Importantly, clinical observations have established that tear film breakup frequently occurs even when tear production is normal and corneal staining is absent \citep{tong2021assessment,tsubota2020new}. This suggests that mechanisms beyond aqueous deficiency are responsible for dry eye syndrome.\\
Multiple mechanisms influence tear film thinning and rupture such as evaporation from the exposed ocular surface \citep{braun2018tear,peng2014evaporation}, non-uniformity of the lipid layer \citep{king2014tear,zhong2019mathematical}, osmotic flux between the tear film and the corneal surface \citep{bruna2014influence}, gravity driven drainage \citep{li2014tear,sahlin1997gravity} and van der Waals forces \citep{craster2009dynamics,sharma1985mechanism}. Among these evaporation, drainage, and osmotic effects are primarily responsible for reducing the film thickness to approximately 0.5 $\mu m$ \citep{braun2018tear,dey2019model}. At such small thicknesses, intermolecular forces become dominant and van der Waals attraction governs the final stages of thinning and rupture. Since these forces arise from interactions between the tear film and the mucin-coated corneal surface, they are highly sensitive to variations in mucin coverage.
 
The  role of membrane-associated mucins in modulating the physicochemical properties of the corneal surface has been getting a lot of attention recently. In healthy eyes, the glycocalyx layer formed by these mucins facilitates uniform spreading and stable adhesion of the tear film. They also reduce friction at the ocular surface \citep{baudouin2019reconsidering,portal2019ocular}. In healthy eyes, this layer facilitates uniform spreading and stable adhesion of the tear film over the corneal surface. However, the glycocalyx may become degraded or spatially non-uniform under pathological conditions associated with DED \citep{stephens2015altered,tong2021assessment,tsubota2020new,gipson2004character,gipson2003role}. Such degradation can arise from several mechanisms such as goblet cell dysfunction or loss \citep{argueso2002decreased}, disruption of the glycocalyx structure \citep{argueso2020disrupted}, increased mechanical friction between the eyelids and the corneal surface \citep{choi2024regional,madl2022mucin} and alterations in the physicochemical properties of mucins themselves \citep{gipson2004character}. \\

From a fluid-mechanical perspective, spatial heterogeneity in mucin coverage generates two distinct effects. First, it introduces variations in the local Hamaker constant which  characterises the strength of van der Waals interactions between the tear film and the corneal surface. Regions of reduced mucin coverage exhibit stronger van der Waals attraction which destabilises the film locally. Second, membrane-associated mucins provide a lubricating interface that facilitates partial slip at the corneal surface \citep{braun2007model}. Spatial variations in mucin coverage therefore induce variations in the local slip length. The slip will be lesser in mucin-deficient regions and higher where mucin coverage is higher. These two effects generate spatial variations in both wettability and interfacial mobility across the corneal surface. This modifies the intermolecular force and capillary pressure gradients that govern film instability \citep{georgiev2019contribution}. \\
 
 Experimental and clinical studies have shown that tear film breakup often occurs in the form of localised dry spots or dimples, which are associated with variations in surface wettability \citep{gipson2004character,tong2021assessment,tsubota2020new}. In particular, tear film is often observed to break at the same spatial locations following successive blinks \citep{tong2021assessment}. This suggests the presence of persistent heterogeneity in surface properties.  Motivated by these observations, we hypothesise that mucin loss occurs in a spatially heterogeneous manner over the corneal surface. Such heterogeneity produces regions of locally enhanced van der Waals attraction and reduced interfacial slip in mucin deficient regions.\\
 
 Thin-film models have been widely used to investigate tear film dynamics, incorporating  capillarity effects, disjoining pressure, evaporation, and surfactant transport. However, most existing models assume spatially uniform wettability and thus a spatially homogeneous base state. For spatially uniform wettability, the capillary pressure gradient, disjoining pressure gradient, and Marangoni stresses all vanish, and the film remains quiescent at steady state. Here, classical normal-mode analysis can then be applied to perform linear stability \citep{burelbach1988nonlinear,dey2019model,craster2009dynamics,zhang2003analysis}. Most existing studies assume a spatially uniform no-slip \citep{li2014tear} or partial-slip boundary condition \citep{dey2019model,zhang2003analysis} at the corneal surface. The membrane-associated mucins form a lubricating glycocalyx layer that facilitates partial slip and reduces interfacial friction. Spatial variations in mucin coverage are therefore expected to produce corresponding variations in slip length along the corneal surface. Such variations modify the local interfacial mobility near the corneal surface and can influence both the stability and rupture dynamics of the tear film. When mucin coverage is spatially heterogeneous, the Hamaker constant varies along the corneal surface. This variation introduces non-zero disjoining pressure gradients even at equilibrium, which must be balanced by capillary forces, resulting in a non-uniform steady-state film thickness. Furthermore, the linearised governing equations then have spatially periodic coefficients. As a result, conventional normal-mode analysis is not valid. \\ 
 A recent study by Choudhury et al \citep{choudhury2021tear} considered spatially varying Hamaker constants arising from mucin heterogeneity. However, the steady-state configuration was assumed to remain uniform, rather than being determined from a balance between capillary and disjoining pressures. Moreover, the linear stability analysis was performed using a normal-mode approach based on a homogeneous system, and the resulting most unstable wavenumber was used to perturb the heterogeneous system. This approach does not account for mode coupling between perturbations induced by spatial heterogeneity. These limitations motivate the development of a theoretical framework that treats spatial heterogeneity consistently in both the base state and the stability analysis. \\

In the present work, we investigate the dynamics of a thin tear film coating a corneal surface with spatially heterogeneous wettability. The heterogeneity is modelled through spatial variation of the Hamaker constant $A_k(x)$  and the slip length $\beta(x)$. We consider two representative functional forms. The first is a piecewise-constant profile which represents sharply localised regions of mucin-rich and mucin-deficient patches on the corneal surface. The second is a smoothly varying sinusoidal profile which represents gradual spatial variations in mucin coverage arising from non-uniform expression or degradation of the glycocalyx layer. Both profiles are assumed to be periodic. This enables the application of Floquet-Bloch theory and discretised eigenvalue analysis  to determine the  linear stability of the steady state. This framework rigorously captures mode coupling induced by spatial heterogeneity which is absent in conventional normal-mode analysis \citep{kuchment2012floquet,pettas2022stability}. The predictions of the linear stability analysis are validated and extended through nonlinear numerical simulations, which reveal the full spatio-temporal evolution of the film and the spatial localisation of rupture. \\
A further motivation for the present model arises from the discrepancy between theoretically predicted and clinically observed tear film rupture times. Previous theoretical studies have reported rupture times in the range of 40-250 s \citep{zhang2003surfactant}, whereas clinical measurements often report breakup times less than 10 s \citep{vanley1977interpretation,norn1969desiccation}. This suggests that existing models may not fully capture all the relevant destabilising mechanisms. In the present study, we propose that spatial heterogeneity in mucin coverage creates localised regions of stronger van der Waals attraction, which can accelerate film thinning and rupture. One of the objectives of this work is therefore to examine whether incorporating such heterogeneity can help bridge the gap between theoretical predictions and clinical observations.\\

The paper is organized as follows. Section \ref{sec:Mathematical Formulation} presents the formulation of the governing equations based on the lubrication approximation. A linear stability analysis is then performed using Floquet- Bloch theory and discretized eigenvalue method in Section \ref{sec:LSA}. We then describe the numerical method for  nonlinear simulations to investigate the effects of mucin heterogeneity on tear film rupture in Section \ref{sec:nonlin}. Section \ref{sec:conclusions} summarizes the key findings, highlights the physiological implications, and discusses possible extensions of the present model.

\section{Mathematical formulation}
\label{sec:Mathematical Formulation}

\subsection {Problem description}
Recent clinical assessments of the tear film report cases of dry eye syndrome in which tear production remains normal and corneal staining is minimal or absent \citep{tong2021assessment,tsubota2020new}. In such situations, dysfunction is often associated with deficiencies or abnormalities in membrane-associated mucins. Experimental observations further suggest that mucin coverage over the corneal surface is not spatially uniform but instead exhibits heterogeneous distributions that may be approximated as periodic variations. Motivated by these observations, we investigate the dynamics and rupture of a thin tear film coating a corneal surface with spatially heterogeneous wettability induced by non-uniform mucin coverage as shown in Figure \ref{fig:1}$(b)$. \\
The corneal epithelium is lined with membrane-associated mucins that form a hydrated glycocalyx layer. This layer promotes wettability and reduces interfacial friction at the ocular surface. Under dry eye conditions, degradation or non-uniform expression of the glycocalyx layer can produce spatial variations in both surface energy and interfacial mobility. Spatial heterogeneity in mucin coverage is incorporated through two surface properties: (i) a spatially varying Hamaker constant $A_k(x)$, representing variations in wettability and intermolecular interactions, and (ii) a spatially varying slip length $\beta(x)$, representing variations in interfacial mobility. The latter accounts for partial slip at the corneal surface due to mucin layer. In the present model, the tear film is treated as a thin Newtonian liquid layer of thickness $h(x,t)$, viscosity $(\mu)$ and surface tension $(\sigma)$.\\
Lipids at the air-tear interface are modelled as an insoluble surfactant that undergoes advection and diffusion along the free surface. The local surfactant concentration is denoted by $\gamma(x,t)$, where the surface tension $\sigma(\gamma)=\sigma_m-S \gamma/\gamma_m$  \citep{zhang2003surfactant}. Here, $\sigma_m$ represents the maximal interfacial tension on the lipid-free interface, $S$ is the maximal spreading pressure and $\gamma_m$ is the maximum lipid concentration. To focus on the role of mucin heterogeneity in tear film dynamics and rupture, we assume isothermal conditions and neglect evaporation and osmotic flux across the ocular surface.

The tear film occupies the region between $0 \leq z \leq h(x,t)$ bounded below by the corneal surface and above by the tear-air interface. The fluid motion within this domain is governed by the incompressible continuity and Navier–Stokes equations,
\begin{equation}
\nabla \cdot \boldsymbol{v} =0,
  \label{eq1}
\end{equation}
\begin{equation}
\rho \left( \frac{\partial \boldsymbol{v}}{\partial t} 
+ \boldsymbol{v} \cdot \nabla \boldsymbol{v} \right)
= - \nabla (p + \phi) + \mu \nabla^{2} \boldsymbol{v}.
  \label{eq2}
\end{equation}
where $\boldsymbol{v}$= $(u,w)$ denotes the velocity field. $u$ and $w$ represent the velocity components along and normal to the corneal surface, respectively, and $p$ is the hydrodynamic pressure. The flow is driven by pressure gradients arising from intermolecular interactions modelled through a potential $\phi$. In the present formulation, this potential accounts for spatially varying wettability (induced by mucin heterogeneity) and is taken as $\dfrac{A_k(x)}{h(x,t)^3}$.\\
At the free surface $z=h(x,t)$, the lipid layer is modelled as an insoluble surfactant whose concentration $\gamma(x,t)$ evolves according to the surface transport equation:
\begin{equation}
\frac{\partial \gamma}{\partial t}
+ \nabla_{s} \cdot (\gamma \boldsymbol{v}_{s})
+ \gamma (\nabla_{s} \cdot \boldsymbol{n})(\boldsymbol{v} \cdot \boldsymbol{n})
= D_{s} \nabla_{s}^{2} \gamma .
\label{eq3}
\end{equation}
where $\boldsymbol{n}$ is the unit normal to the free surface $z=h(x,t)$ and $\nabla_{s}=(\boldsymbol{I}-\boldsymbol{nn}) \cdot \nabla$ is the surface gradient operator, and  $\boldsymbol{v}_s=\boldsymbol{v}-\boldsymbol{nn} \cdot \boldsymbol{v}$  is the tangential surface velocity. The parameter $D_s$ denotes the surface diffusivity of the lipids.\\
At the corneal surface, $z=0$, we impose a no-penetration condition together with a Navier slip boundary condition characterised by a spatially varying slip length $\beta(x)$.
\begin{equation}
\boldsymbol{v}_{t} = \beta(x) \, (\boldsymbol{n}_{c} \cdot \nabla \boldsymbol{v}_{t}),
\qquad
\boldsymbol{n}_{c} \cdot \boldsymbol{v} = 0 .
\label{eq4}
\end{equation}
where, $\boldsymbol{v}_t=\boldsymbol{t}_c \cdot \boldsymbol{v}$ denotes the tangential velocity along the corneal surface. Here, $\boldsymbol{t}_c$ and $\boldsymbol{n}_c$ are the unit tangent and normal vectors respectively defined on the corneal surface $z=0$. The expressions for these vectors are provided in the supplementary material (Section S.1).\\
At the free surface $z=h(x,t)$, the balance of tangential stresses is given by,
\begin{equation}
\boldsymbol{n} \cdot \boldsymbol{\tau} \cdot \boldsymbol{t} = \nabla_s \sigma \cdot \boldsymbol{t}
\label{eq5}
\end{equation}
where, $\boldsymbol{\tau}$ is the deviatoric stress tensor and $\boldsymbol{t}$ is a unit tangent vector to the interface $z=h(x,t)$. This condition captures  the Marangoni stress arising from surface tension gradients along the interface. The normal stress balance gives the relationship between the surface tension, and the pressure jump across the tear-air interface. 
\begin{equation}
\boldsymbol{n} \cdot \boldsymbol{\tau} \cdot \boldsymbol{n} =(p-p_\text{atm})-\sigma(\nabla \cdot \boldsymbol{n})
\label{eq6}
\end{equation}
where, $p$ denotes the pressure within the liquid film and $p_\text{atm}$ is the pressure above the tear-air interface. The term $ \nabla \cdot \boldsymbol{n}$ represents the curvature of the interface. The free surface  $z=h(t,x)$ evolves according to the kinematic condition,
\begin{equation}
\frac{\partial h}{\partial t}
+ u \frac{\partial h}{\partial x}
= w .
\label{eq7}
\end{equation}
Explicit expressions for the stress components and other variables are provided in the section S.1 of the supplementary material. The physical parameters characterizing the tear film and corneal surface are summarized in Table \ref{tab:1}. 
\subsection{Modeling mucin heterogeneity}
The  heterogeneity of  mucin coverage is modelled using two periodic representations of the Hamaker constant $A_k (x)$.

\subsubsection{Periodic step variation}
Here, the corneal surface consists of alternating mucin-rich and mucin-deficient regions with the concentration being constant in each region. This is represented by a periodically repeating step variation in the Hamaker constant,
\begin{equation}
A_k(x)=
\begin{cases}
A_{k1}, & \text{for} \quad 0 \le x < fL, \\[4pt]
A_{k2}, & \text{for} \quad fL \le x < L .
\end{cases}
\label{eq8}
\end{equation}
with periodic extension in the $x$-direction. Here, $f$ represents the fraction of domain which is mucin rich. $ A_{k1}$ and $A_{k2}$ denote the Hamaker constants corresponding to mucin-rich (healthy eye) and mucin-deficient regions, respectively with $A_{k2}>A_{k1}$. The larger Hamaker constant in mucin-deficient regions represents stronger effective intermolecular attraction arising from the loss of the hydrated glycocalyx layer. This represents situations in which mucin degradation occurs in sharply localised patches due to epithelial damage, inflammation, or mechanical abrasion.
Mucin heterogeneity also modifies interfacial mobility at the corneal surface. The glycocalyx layer provides a lubricating interface that facilitates partial slip over the corneal surface. In mucin-deficient regions, this lubricating effect is lesser resulting in reduced slip. To account for this behaviour, the slip length is prescribed as
\begin{equation}
\beta(x)=
\begin{cases}
\beta_{1}, & \text{for} \quad 0 \le x < fL, \\[4pt]
\beta_{2}, & \text{for} \quad fL \le x < L .
\end{cases}
\label{eq9}
\end{equation}
Here, $\beta_1$  and $\beta_2$ represent the slip lengths in mucin-rich and mucin-deficient regions, respectively. In the present study, the limiting case $\beta_2=0$ is considered. This corresponds to a no-slip condition associated with exposure of the underlying epithelial surface.

\subsubsection{Smooth sinusoidal variation}
In the second representation, the Hamaker constant $A_k (x)$  varies smoothly and periodically along the corneal surface. The Hamaker constant is prescribed here  as
\begin{equation}
A_k(x)=A_{k0}\left(1+\epsilon \sin\left(\frac{k_s x}{L}\right)\right)
\label{eq10}
\end{equation}
where $A_{k0}$  denotes the mean Hamaker constant, $\epsilon$ represents the amplitude of the spatial variation and $k_s=2\pi$. This form ensures a continuous modulation of intermolecular interactions with wavelength $L$.
The corresponding variation in slip length is taken as
\begin{equation}
\beta(x)=\beta_{k0}\left(1-\epsilon \sin\left(\frac{k_s x}{L}\right)\right)
\label{eq11}
\end{equation}
This choice assures that regions with larger $A_k (x)$, are associated with smaller slip lengths. 

\subsection {Nondimensionalization}
The governing equations and the boundary conditions are nondimensionalized using the following characteristic scales,\\
$$x_c =L, z_c=H, A_{kc}=A_{k1},,u_c=U=\frac{A_{k1}}{6 \pi \mu H L},w_c =\epsilon U, t_c = \frac{L}{U}$$
$$p_c=\frac{A_{k1}}{6 \pi H^3},h_c=H,\beta_c=H,\gamma_c=\gamma_m,\sigma_c=\sigma_m$$
Here, $L$ and $H$ denote the characteristic length scales in $x$ and $z$ directions respectively.  Since van der Waals interaction is the dominant mechanism driving tear film rupture, the characteristic velocity scale $U$ is determined by balancing viscous forces with van der Waals forces. The Hamaker constant for a healthy mucin $A_{k1}$  is chosen as the characteristics scale. The pressure scale is chosen based on the characteristic magnitude of the disjoining pressure, $\frac{A_{k1}}{6 \pi H^3}$. Following nondimensionalisation, the spatial variations in  $A_k (x)$ and slip length $\beta(x)$ for periodic step variation are prescribed as:
\begin{equation}
A_k(x)=
\begin{cases}
1, & 0 \le x < f, \\[4pt]
A_r, & f \le x < 1 ,
\end{cases}
\qquad 
\beta(x)=
\begin{cases}
\beta_0, & 0 \le x < f, \\[4pt]
0, & f \le x < 1 .
\end{cases}
\label{eq12}
\end{equation}
Here, $A_r= \frac{A_{k2}}{A_{k1}}>1$ represents the increased intermolecular attraction in mucin-deficient regions, while $\beta(x)=0$ corresponds to loss of the lubricating glycocalyx layer and $\beta_0$ denotes the dimensionless slip length. Direct experimental measurements of $A_r$ for tear films over mucin-depleted corneal surfaces are currently unavailable. However, Choudhary \textit{et al}  estimated this quantity theoretically and reported that $A_r$ may vary over the range 1.16-46.5 \citep{choudhury2021tear}.  To model sinusoidal spatial variations in mucin coverage, $A_k (x)$ and $\beta(x)$ are given as,
\begin{equation}
    A_k(x) = A_{0} \left(1 + \epsilon \sin(k_s x)\right) \qquad \text{and} \qquad  \beta(x) = \beta_{0} \left(1 - \epsilon \sin(k_s x)\right)
    \label{eq13}
\end{equation}
$A_0=\frac{A_{k0}}{A_{k1}}$ represents the dimensionless mean Hamaker constant relative to that of a healthy eye and  $\beta_{0}$ denotes the dimensionless mean slip length.

\begin{table}
\begin{center}
\def~{\hphantom{0}}
\begin{tabular}{lll}
\textbf{Symbol} & \textbf{Description} & \textbf{Value (Reference)} \\[5pt]
$H$ & Characteristic thickness & $0.5\times10^{-6}\ \mathrm{m}$ \citep{dey2019model} \\
$L$ & Characteristic length & $1.5\times10^{-4}\ \mathrm{m}$ \citep{luke2021parameter} \\
$\rho$ & Density of tear film & $1000\ \mathrm{kg\,m^{-3}}$ \citep{deng2014heat} \\
$\mu$ & Viscosity of tear film & $1.3\times10^{-3}\ \mathrm{Pa\,s}$ \citep{tiffany1991viscosity} \\
$\sigma_m$ & Maximum interfacial tension & $4.5\times10^{-2}\ \mathrm{N\,m^{-1}}$ \citep{nagyova1999components} \\
$\beta_1$ & Slip coefficient & $3.5\times10^{-7}\ \mathrm{N\,m^{-1}}$ \citep{zhang2003analysis} \\
$S$ & Maximum spreading pressure & $7.5\times10^{-8}\ \mathrm{N\,m^{-1}}$ \citep{zhang2003analysis} \\
$\gamma_m$ & Maximum lipid concentration & $4\times10^{-7}\ \mathrm{mol\,m^{-2}}$ \citep{bruna2014influence} \\
$A_{k1}$ & Unretarded Hamaker constant & $6 \pi \times 3.5\times10^{-19}\ \mathrm{Pa\,m^{3}}$ \citep{winter2010model} \\
$D_s$ & Surface diffusivity & $10^{-11}\ \mathrm{m^{2}\,s^{-1}}$ \citep{adalsteinsson2000lipid} \\
\end{tabular}
\caption{Physical parameters used in the mathematical model and their corresponding reference sources.}
\label{tab:1}
\end{center}
\end{table}
\subsection {Lubrication approximation}
We define $ \delta =\frac{H}{L} \ll 1 $ and exploiting this, we apply the lubrication approximation to simplify the governing equations. Under this framework, only leading-order terms are retained, while terms of order $\delta$ or smaller are neglected. The resulting nondimensionalized governing equations are derived in the Supplementary Material (Section S.2). The dimensionless variables are written without the superscript  *.\\
At the leading order O(1), we obtain
\begin{equation}
\frac{\partial u}{\partial x}+\frac{\partial w}{\partial z}=0,
\label{eq14}
\end{equation}
\begin{equation}
-\frac{\partial p}{\partial x}-\frac{\partial \phi}{\partial x} + \frac{\partial^2 u}{\partial z^2}=0,
\label{eq15}
\end{equation}
\begin{equation}
-\frac{\partial p}{\partial z}=0.
\label{eq16}
\end{equation}
 The surfactant transport equation (\ref{eq3})  in dimensionless form is given as,
\begin{equation}
\frac{\partial \gamma}{\partial t}
+ \frac{\partial}{\partial x}\!\left( u_{s} \gamma \right)
= \frac{1}{Pe_{s}}
\frac{\partial^{2} \gamma}{\partial x^{2}} .
\label{eq17}
\end{equation}
Here, $u_s$ denotes the $x$-component of surface velocity and $Pe_{s}=UL/D_s$  is the Peclet number for mucin diffusion. This is subject to  the boundary conditions,\\
At $z=0$, 
\begin{equation}
u=\beta(x) \frac{\partial u}{\partial z},                        
\label{eq18}
\end{equation}
and 
\begin{equation}
w=0.                        
\label{eq19}
\end{equation}
At $z=h(x)$, 
\begin{equation}
p=-C \frac{\partial^2 h}{\partial x^2},                        
\label{eq20}
\end{equation}
and 
\begin{equation}
\frac{\partial u}{\partial z} =-M \frac{\partial \gamma}{\partial x} . 
\label{eq21}
\end{equation}
Here,  $C=\frac{\delta^3 \sigma_m}{\mu U}$ is the reduced Capillary number and $M= \frac{SH}{\mu U L }$ is the Marangoni number. Integrating the  continuity equation in the $z$-direction from $z=0$ to  $h(x,t)$ , we obtain\\
\begin{equation}
\int_{0}^{h(x,t)} 
\frac{\partial u}{\partial x}\, \mathrm{d}z
+ \left. w \right|_{z = h(x,t)}
- \left. w \right|_{z = 0}
= 0 .  
\label{eq22}
\end{equation}
Using Leibniz's rule of integration and the kinematic boundary condition yields ,
\begin{equation}
\dfrac{\partial h}{\partial t}+ \dfrac{\partial}{\partial x}\int_{0}^{h(x,t)} 
u\, \mathrm{d}z
- \left. w \right|_{z = 0}
= 0 .  
\label{eq23}
\end{equation}

 The dependent variable $u$ and $w$ are functions of time and spatial coordinates $x$ and $z$, while $p,\phi,h$ and $\gamma$ depend only on time and $x$  coordinate.\\
 Since the free surface of the tear film is deformable, the physical domain is time dependent and given by  $z \in [0,h(x,t)]$. To avoid solving the governing equations on a time dependent domain, we introduce a coordinate transformation that maps the evolving film region onto a fixed rectangular domain. For this, we define a transformed vertical coordinate,
 \begin{equation}
\zeta = \frac{z}{h(x,t)}.
\label{eq24}
\end{equation}
This transformation maps the physical tear-film region onto a fixed rectangular computational domain defined by $x\in[0,1]$ and $\zeta \in [0,1]$. All governing equations and boundary conditions are subsequently expressed in the $(x,\zeta)$ domain. The transformed continuity equation is 
\begin{equation}
\frac{\partial u}{\partial x} 
- \frac{\zeta  h' }{h}
\frac{\partial u}{\partial \zeta}
+ \frac{1}{h} \,
\frac{\partial w}{\partial \zeta}
= 0 .
\label{eq25}
\end{equation}
The primes $(')$ denote the derivative with respect to $x$. The pressure within the tear film is governed by capillary effects associated with the curvature of the free surface given by equation (\ref{eq20}). Substituting this expression into the $x$-momentum equation (\ref{eq15}) we obtain 
\begin{equation}
C \frac{\partial^3 h}{\partial x^3} +\frac{3A_k(x)h'}{h^4} - \frac{A_k'(x)}{h^3}+\frac{1}{h^2} \frac{\partial^2 u}{\partial \zeta^2}=0.
\label{eq26}
\end{equation}
The surfactant transport equation and kinematic boundary conditions remain unchanged, as they are defined at $\zeta=1$. The boundary conditions (\ref{eq18}), (\ref{eq19}) and (\ref{eq21}) are,
\begin{equation}
u=\frac{\beta(x)}{h} \frac{\partial u}{\partial \zeta}  \quad \text{at} \quad \zeta =0,          
\label{eq27}
\end{equation}
\begin{equation}
w=0    \quad \text{at} \quad \zeta =0,            
\label{eq28}
\end{equation}
and 
\begin{equation}
\frac{1}{h}\frac{\partial u}{\partial \zeta} =-M \frac{\partial \gamma}{\partial x} \quad \text{at} \quad \zeta =1.
\label{eq29}
\end{equation}
Integrating equation (\ref{eq26}) twice with respect to $\zeta$ yields:
\begin{equation}
u=c_1 (x)  \frac{\zeta^2}{2}+c_2 (x)  \zeta+c_3 (x),
\label{eq30}
\end{equation}
where, $c_1(x)=- h^2\left(\frac{3 A_k h'}{h ^4}+ C h'''\right)$. Using equation (\ref{eq27}), $c_3=\frac{\beta(x) c_2 (x)}{h}$. The tangential stress balance equation (\ref{eq29}) implies  
\begin{equation}
c_2 (x)=-M\gamma' h-c_1 (x).
\label{eq31}
\end{equation}
Substituting the values of $c_1 (x)$, $c_2 (x)$ and $c_3 (x)$ in equation (\ref{eq30}) yields,
\begin{equation}
\begin{aligned}
u
&=
-\left[
-h^2\left(
\frac{3A_k(x)h'}{h^4}
+Ch'''
\right)
+\frac{A_k'(x)}{h}
\right]
\frac{\zeta^2}{2}
\\[6pt]
&\quad
+\left[
-M\gamma' h
+h^2\left(
\frac{3A_k(x)h'}{h^4}
+Ch'''
\right)
-\frac{A_k'(x)}{h}
\right]
\left(
\frac{\beta(x)}{h}
+\zeta
\right)
\end{aligned}
\label{eq33}
\end{equation}
From continuity equation (\ref{eq25}) and equation (\ref{eq33}), we obtain 
\begin{equation}
\begin{aligned}
w
&=
-\frac{1}{6}\zeta^3 h\, c_1'(x)
-\frac{1}{2}\zeta^2 h\, c_2'(x)
\\[6pt]
&\quad
+\zeta\left(
-\beta(x)c_2'(x)
-c_2(x)\beta'(x)
+\frac{c_2(x)\beta(x)h'}{h}
\right)
\end{aligned}
\end{equation}

Substituting the expression of $u$ into equation (\ref{eq17}) and equation (\ref{eq23}), we obtain the evolution equations for film thickness and the lipid concentration, 
\begin{equation}
\begin{aligned}
\frac{\partial h}{\partial t}
&=
A_k''(x)\left(\frac{\beta(x)}{h}+\frac{1}{3}\right)
+A_k'(x)
\left(
\frac{
\beta'(x)h
-h'(h+4\beta(x))
}{h^2}
\right)
\\
&\quad
+A_k(x)
\left(
\frac{
-3\beta'(x)hh'
+{h'}^2(h+6\beta(x))
-hh''(h+3\beta(x))
}{h^3}
\right)
\\[6pt]
&\quad
+C\Bigg(
-h'''h\Big(h'(h+2\beta(x))+\beta'(x)h\Big)
-\frac{1}{3}h''''h^2(h+3\beta(x))
\Bigg)
\\
&\quad
+M\Bigg(
h'\gamma'(h+\beta(x))
+\beta'(x)h\gamma'
+\frac{1}{2}h\gamma''(h+2\beta(x))
\Bigg)
\end{aligned}
\label{eq35}
\end{equation}
and,
\begin{equation}
\begin{aligned}
\frac{\partial \gamma}{\partial t}
&=
A_k''(x)
\left(
\frac{\gamma(h+2\beta(x))}{2h^2}
\right)
+A_k'(x)
\left(
\frac{
-2\gamma h'(2h+5\beta(x))
+2\gamma h\beta'(x)
+h\gamma'(h+2\beta(x))
}{2h^3}
\right)
\\[6pt]
&\quad
+A_k(x)
\Bigg(
-\frac{3}{2h^4}
\Big[
\gamma hh''(h+2\beta(x))
+hh'\big(\gamma'(h+2\beta(x))
+2\gamma\beta'(x)\big)
-2\gamma(h')^2(h+3\beta(x))
\Big]
\Bigg)
\\[6pt]
&\quad
+C\Bigg(
-\gamma hh^{(3)}\beta'(x)
-\frac{1}{2}
h\big(\gamma h^{(4)}+h^{(3)}\gamma'\big)
(h+2\beta(x))
-\gamma h^{(3)}h'(h+\beta(x))
\Bigg)
\\[6pt]
&\quad
+M\Bigg(
\gamma\gamma' h'
+\big(\gamma\gamma''+(\gamma')^2\big)
(h+\beta(x))
+\gamma\gamma'\beta'(x)
\Bigg)
+\frac{\gamma''}{Pe_s}
\end{aligned}
\label{eq36}
\end{equation}

Equations (\ref{eq35}-\ref{eq36}) govern the nonlinear evolution of the film thickness and surfactant concentration over a corneal surface with spatially varying wettability and interfacial slip. In the limiting case where $A_k (x)$ and $\beta(x)$ are constants, the governing equations reduce to the classical evolution equations for a thin film with insoluble surfactant on a homogeneous substrate. In this limit, the present formulation recovers the model of Zhang \textit{et al} \cite{zhang2003surfactant}, thereby demonstrating the mathematical consistency of the derivation. We next  compute the steady-state solution of the system.

\subsection{Steady states}
We consider the steady state, where  the tear film remains stationary and therefore $u=w=0$. The base-state lipid concentration is spatially uniform as implied by the tangential stress balance (equation (\ref{eq29})). Consequently, no surface tension gradients are present. Marangoni stresses are therefore absent at equilibrium. At steady state, the film profile is determined by a balance between capillary forces and intermolecular (van der Waals) forces, which act even in the absence of fluid motion. The steady-state equation is therefore obtained from the $x$-momentum balance as
\begin{equation}
    Ch'''(x) + \frac{3 A_k(x) h'(x)}{h^4}-\frac{A'_k(x)}{h^3}=0,
\label{eq37}
\end{equation}
which is a nonlinear third-order ordinary differential equation for the film thickness $h(x)$. This equation represents the balance between the capillary pressure gradient arising from interfacial curvature and the disjoining pressure induced by spatially varying intermolecular interactions. \\
For a step change in Hamaker constant, the uniform film thickness $h_s (x)=1$ satisfies equation (\ref{eq37}) \citep{kargupta2000instability,kargupta2002dewetting}. Since $A_k (x)$ is piecewise constant, its spatial derivatives vanish within each region. Both the capillary pressure gradient and the disjoining pressure gradient are identically zero. As a result, no flow is induced and the tear film remains spatially uniform at steady state. However, there is a singularity at $x=f$ where there is a sharp change in the disjoining pressure. In particular, the steady state solution is independent of the magnitude of $A_r$. In contrast, when $A_k (x)$ varies smoothly  (equation (\ref{eq13})), the disjoining-pressure gradient becomes non-zero throughout the domain. To maintain equilibrium, this must be balanced by a corresponding capillary pressure gradient. This must result in a spatially non-uniform steady-state film profile.  

\begin{figure}
  \centering
  \includegraphics[width=1\textwidth]{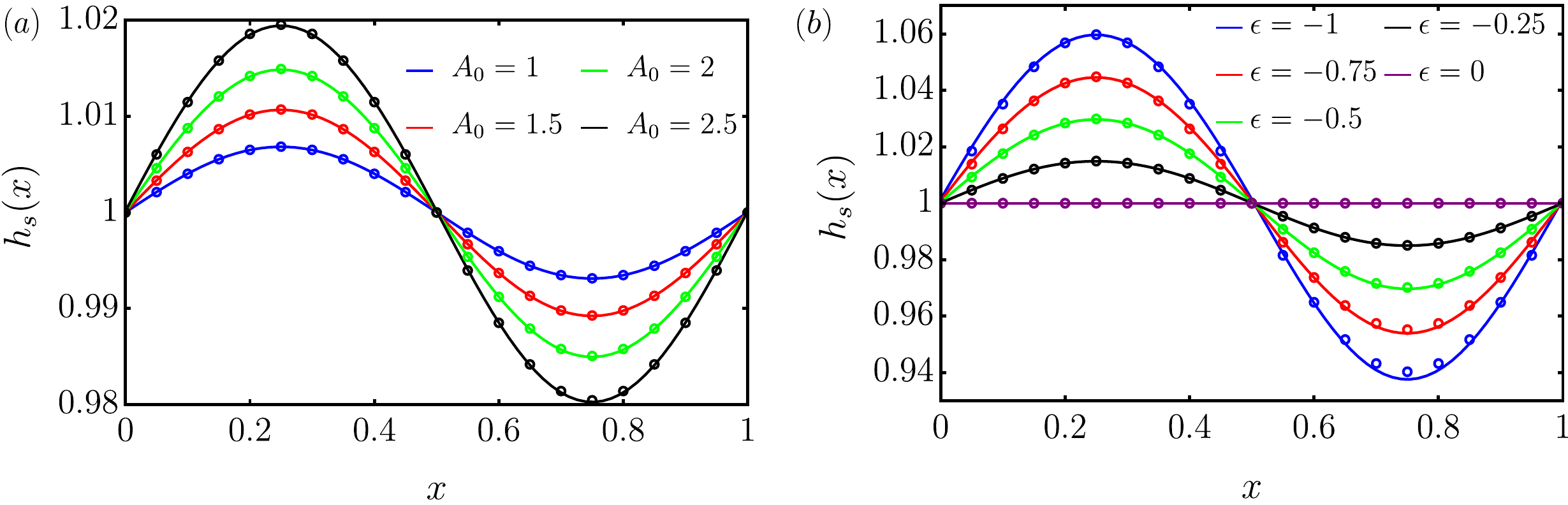}
  \caption{ Steady-state film profiles for a sinusoidally varying $A_k (x)$ exhibiting a spatially non-uniform film profile $(a)$ Steady state film thickness for different $A_0$ for $\epsilon=-0.25$ $(b)$ Steady state for different $\epsilon$ for $A_0=2$. The other parameter is $C=1$.}
  \label{fig:2}
\end{figure}

\subsubsection{Asymptotic analysis in the weak mucin heterogeneity limit for sinusoidal variation}
Although the governing equation is nonlinear, an analytical solution can be obtained in the limit of weak heterogeneity, $\epsilon \ll 1$. The film thickness $h(x)$ is expanded as a regular perturbation series,
\begin{equation}
   h(x) = h_0(x) + \epsilon h_1(x) +\epsilon^2 h_2(x) +...
\label{eq38}
\end{equation}
Here, $h_0 (x)$ denotes the uniform base thickness corresponding to a homogeneous corneal surface, while the higher-order terms represent modifications induced by mucin heterogeneity. At the leading order, $O(1)$, the governing equation reduces to
\begin{equation}
 Ch_0'''(x) + \frac{3 A_k(x)}{h_0^4} h_0'(x)=0
\label{eq39}
\end{equation}
This is subject to periodic boundary conditions and admits only spatially uniform solutions. Without loss of generality, the mean film thickness is normalised such that $h_0 (x)=h_{ss} =1.$ 
At $O(\epsilon)$, the governing equation (\ref{eq37}) becomes
\begin{equation}
C h_1'''(x)
+\frac{3A_0}{h_{ss}^4}h_1'(x)
-\frac{A_0 k_s \cos(k_s x)}{h_0^3}
=0
\label{eq40}
\end{equation}
This equation describes the linear response of the steady film profile to weak spatial heterogeneity in the Hamaker constant. Imposing periodic boundary conditions yields $h_1(x)=
\frac{A_0}{3A_0-Ck_s^2}\sin(k_s x)$. Reconstructing the solution gives the analytical expression for the steady-state film profile, 
\begin{equation}
h_s(x) = 1 + \frac{A_0 \epsilon}{3A_0-Ck_s^2} \sin(k_s x)
\label{eq41}
\end{equation}
 The amplitude of the steady-state deformation depends on the reduced capillary number $C$, $k_s$, and the mean Hamaker constant $A_0$.  

\subsubsection{Numerical computation of steady-state solutions}
The analytical solution (equation (\ref{eq41})) is not valid when $\epsilon$ is not asymptotically small. For large $\epsilon$, the steady-state equation (\ref{eq37}) is solved numerically using a Fourier spectral discretisation \citep{trefethen2000spectral,weideman2000matlab} on the periodic domain $x \in [0,1]$. The periodic domain $x \in [0,1]$ is discretised using $N_p=73$ uniformly spaced collocation points. Spatial derivatives are evaluated using Fourier differentiation matrices constructed from the discrete Fourier transform. This provides spectral accuracy up to third-order derivatives. \\
Let $h_i$ denote the discrete approximation of the film thickness at the $i$-th collocation point. The discrete solution is then assembled into the vector $ h \ \epsilon \ R^{N_p }$. Under periodic boundary conditions, the steady-state equation admits a family of solutions corresponding to different mean film thicknesses. To obtain a unique solution, conservation of liquid volume is imposed through the constraint, 
\begin{equation}
 \int_{0}^{1}  h(x)  \, dx = 1.   
\end{equation}

The resulting system of nonlinear algebraic equations is solved using a Newton-type iterative method implemented via \textit{FindRoot} in \textsc{Mathematica}. The iteration is initialized with a uniform film profile $h_i=1$, and convergence is assumed to be attained when the residual norm falls below $10^{-10}$. \\
Figure \ref{fig:2}$(a)$ illustrates the steady-state film profile for a sinusoidally varying Hamaker constant. Unlike the piecewise-constant case, the spatial variation of $A_k (x)$ introduces non-zero gradients in both capillary and disjoining pressure terms. At steady state, these competing effects balance each other. This leads to a spatially non-uniform film profile, as shown in Figure \ref{fig:2}$(a)$. The amplitude of the steady-state deformation increases with increasing $A_0$. This is physically consistent and arises from the stronger intermolecular attraction between the corneal surface and the tear-air interface with increasing $A_0$. The substrate in the right half, $x \geq0.5$, exhibits stronger intermolecular attraction than the left half. Consequently, the steady-state film thickness is reduced in this region compared to the left side. The solid lines in Figure \ref{fig:2}$(a)$ represent the numerical solution, while the open circles denote the analytical prediction from equation (\ref{eq41}). The solutions obtained using the two approaches show excellent agreement.  Figure \ref{fig:2}$(b)$ shows the steady-state film configuration for different values of $\epsilon$. The amplitude of the film deformation at the steady state increases with increasing $|\epsilon|$. In the limiting case $\epsilon=0$, the Hamaker constant becomes uniform and the steady-state solution reduces to a flat film i.e. $h_s(x)=1$.\\
\vspace{-2em}
\section{Linear stability analysis}
\label{sec:LSA}

\subsection{Floquet-Bloch Theory}
To examine the linear stability of the steady state, we introduce small perturbations to the film thickness, lipid concentration, and velocity components of the form
\begin{equation}
\begin{aligned}
h(x,t) &= h_s(x) + \eta \ h_1(x,t),\\
\gamma(x,t) &= \gamma_s + \eta \ \gamma_1(x,t),\\
u(x,t) &=  \eta \ u_1(x,t),\\
w(x,t) &=  \eta \ w_1(x,t).
\end{aligned}
\label{eq42}
\end{equation}

where $ \eta \ll 1$ and the variables $h_1,\gamma_1,u_1,w_1$ represent small perturbations to the base state. Substituting expressions (\ref{eq42}) into the governing equations (\ref{eq14}-\ref{eq23}) and retaining terms up to $O(\eta)$ yields a linearized system describing the evolution of these perturbations. The complete set of resulting linearized equations and boundary conditions is presented in Section S.3 of  the Supplementary Material.\\ 
By solving the linearized momentum equations, the velocity perturbations $u_1$  and $w_1$ can be written explicitly in terms of the film thickness and surfactant perturbations, $h_1$ and $\gamma_1$. These expressions are provided in Appendix \ref{App:A}. Substituting the resulting velocity fields into the linearized kinematic condition and the surfactant transport equation leads to a system of coupled evolution equations governing the perturbations.     \begin{subequations}\label{eq43}
\begin{align}
\frac{\partial h_1}{\partial t}
&=
P_1(x) h_1
+ Q_1(x) h_1'
+ R_1(x) h_1''
+ S_1(x) h_1'''
+ T_1(x) h_1''''
+ U_1(x) \gamma_1'
+ V_1(x) \gamma_1''
\\
\frac{\partial \gamma_1}{\partial t}
&=
P_2(x) h_1
+ Q_2(x) h_1'
+ R_2(x) h_1''
+ S_2(x) h_1'''
+ T_2(x) h_1''''
+ U_2(x) \gamma_1'
+ V_2(x) \gamma_1''
\end{align}
\end{subequations}

The coefficient functions $ P_i (x),Q_i (x),R_i (x),S_i (x),T_i (x),U_i (x)$ and $V_i (x)$ for $i=1,2$ are spatially periodic. This periodicity arises from the periodic variation of the Hamaker constant $A_k (x)$, the slip length $\beta(x)$, and the steady-state film thickness $h_s (x)$. Explicit expressions for these coefficients are provided in Appendix \ref{App:A}. Since these coefficients are spatially dependent, normal mode linear stability analysis cannot be used. Hence, we employ Floquet-Bloch theory to determine the stability characteristics of the system. Accordingly, we seek the solution to the perturbations in Bloch form as \citep{kuchment2012floquet,ajaev2013application},
\begin{equation}
h_1 = e^{\sigma t} e^{\alpha x} \phi(x),
\label{eq45}
\end{equation}
and
\begin{equation}
\gamma_1 = e^{\sigma t} e^{\alpha x} \psi(x).
\label{eq46}
\end{equation}
where $\sigma$ is the temporal growth rate. $\phi(x)$ and $\psi(x)$ are the spatial eigenfunctions. Exploiting the periodicity of the coefficients, we find that $\alpha=iq$, where $q \in [-\pi,\pi]$ defines the first Brillouin zone. This has been derived in Appendix \ref{App:B}. Restricting attention to this interval is sufficient to characterize the full stability spectrum. 
The corresponding spatial derivatives are  given by

\begin{equation}
\begin{aligned}
\frac{d^{n} h_1}{dx^{n}}
&=
e^{\sigma t} e^{i q x}
\left( \frac{d}{dx} + i q \right)^{n} \phi(x),\\
\frac{d^{n} \gamma_1}{dx^{n}}
&=
e^{\sigma t} e^{i q x}
\left( \frac{d}{dx} + i q \right)^{n} \psi(x).  
\end{aligned}
\label{eq47}
\end{equation}

The exponent $n$ in equation (\ref{eq47}) denotes the order of the derivative and $\frac{d}{dx}$ represents the first order derivative. The domain $[0,1]$ is discretised using a uniform grid of $N_p$ collocation points defined by,
\begin{equation}
x_j = \frac{j}{N_p}, 
\qquad
j = 0,1,2,\ldots, N_p - 1.
\label{eq49}
\end{equation}
The coefficients $ P_i (x),Q_i (x),R_i (x),S_i (x),T_i (x),U_i (x)$ and $V_i (x)$ are evaluated at these collocation points. The periodic functions  $\phi(x)$ and  $\psi(x)$ are discretized using $N_p$ Fourier nodes. Spatial derivatives are approximated using Fourier differentiation matrices \citep{weideman2000matlab}. The first-order derivative is represented by the differentiation matrix $\mathbb{D}_1$ while higher-order derivatives are obtained using matrix products,
\begin{equation}
\mathbb{D}_2 = \mathbb{D}_1\cdot \mathbb{D}_1, \mathbb{D}_3 = \mathbb{D}_2\cdot \mathbb{D}_1 \quad \text{and} \quad \mathbb{D}_4 = \mathbb{D}_2\cdot \mathbb{D}_2
\label{eq50}
\end{equation}

Substituting the Bloch ansatz into equations (\ref{eq43}) and replacing differential operators with their discrete counterparts, yields an eigenvalue problem for the growth rate $\sigma$ parameterized by the Bloch wavenumber $q$. The resulting system can be written in matrix form as,
\begin{equation}
\sigma \mathbf{Z} =
\begin{pmatrix}
A_{h\phi}(q) & A_{h\psi}(q) \\
A_{\gamma\phi}(q) & A_{\gamma\psi}(q)
\end{pmatrix}
\mathbf{Z}
\label{eq51}
\end{equation}
Here, $\mathbf{Z}=\left[\phi_1,\, \phi_2,\, \ldots,\, \phi_{N_p}, \psi_1,\, \psi_2,\, \ldots,\, \psi_{N_p}\right]^{T}$ is the vector of unknown variables evaluated at the collocation points. The matrices $A_{h\phi}(q)$ and $A_{h\psi}(q)$ (each of size $N_p \times N_p$) arise from the coefficients of $\phi$ and $\psi$ in equation (\ref{eq43}a). Similarly, the matrices $A_{\gamma\phi}(q)$ and $A_{\gamma\psi}(q)$ are obtained from the coefficients of $\phi$ and $\psi$ in equation (\ref{eq43}b). The dependence on the Bloch wavenumber $q$ enters through the modified differential operator $\left(\dfrac{d}{dx}+iq\right)$. For each prescribed value of $q$, the resulting problem (equation \ref{eq51}) admits a discrete spectrum of eigenvalues. The stability of the system is determined by the dominant eigenvalue $\sigma_m=\max\big(\mathrm{Re}(\sigma)\big)$. The dispersion relation $\sigma_m (q)$ is obtained by evaluating the leading growth rate over the first Brillouin zone. This framework generalizes classical normal-mode stability analysis by explicitly accounting for spatially periodic coefficients.
\subsection{Discretized eigenvalue approach}
To validate the dispersion curves from Floquet theory, we analyse the coupled linear system (\ref{eq43}) using  an alternative  approach. This is based on a Fourier-Galerkin expansion to compute discrete eigenvalues. Since the substrate pattern is periodic, we consider an extended computational domain of length $(L_e >> l)$ consisting of $N_g$ repeated unit cells such that $L_e=l \times N_g$. Here $l=1$ is the fundamental wavelength of substrate pattern. The perturbations $h_1$ and $\gamma_1$ are expanded as Fourier series over the extended domain $L_e$, 

\begin{equation}
\begin{aligned}
h_1
&=
\sum_{n=1}^{\infty}
\left(
h_n e^{i q_n x}
+
\bar{h}_n e^{-i q_n x}
\right),\\
\gamma_1
&=
\sum_{n=1}^{\infty}
\left(
\gamma_n e^{i q_n x}
+
\bar{\gamma}_n e^{-i q_n x}
\right).
\label{eq52d}
\end{aligned}
\end{equation}

Here, $q_n=\frac{2 \pi n}{L_e}$   are the discrete wavenumbers. $\bar{h}_n$, $\bar{\gamma}_n$ denotes the complex conjugate of $h_n$ and $\gamma_n$ respectively. 
The spatially periodic coefficient $P_i (x)$ for $i=1,2$ are similarly expanded as,

\begin{equation}
P_i(x)
=
\sum_{j=0}^{\infty}
\left(
P_{ij} e^{i q_j x}
+
\text{c.c.}
\right)
\label{eq54d}
\end{equation}

where $q_j=\frac{2 \pi j}{L_e}N_g=\frac{2 \pi j}{l}$ and the Fourier coefficients $(P_{ij})$ are computed by,
\begin{equation}
P_{ij}
=
 \int_{0}^{1}
P_i(x)\, e^{-i q_j x}\, dx
\label{eq55d}
\end{equation}
All remaining periodic coefficient $Q_i (x),R_i (x),S_i (x),T_i (x),U_i (x)$ and $V_i (x)$ are expanded analogously as equation (\ref{eq54d}). These coefficients are evaluated at the steady-state solution $h_s (x)$ which is a smooth function of $x$. Hence, the Fourier coefficients decay rapidly with $j$. This ensures fast convergence of the truncated expansions. We substitute the Fourier expansions (\ref{eq52d}) into the linearized equations (\ref{eq43}). Multiplying by $e^{-iq_m x}$, and integrating over the full computational domain $[0,L_e]$, we exploit the orthogonality relation
\begin{equation}
    \frac{1}{L_e}\int_{0}^{L_e}
 e^{-i( q_a - q_b )x}\, dx =\delta_{ab}
\label{eq56d}
\end{equation}
when we project the $n^{th}$ Fourier mode onto the $m^{th}$ Fourier mode. After projection, the non-zero contributions to row $m$ arise only from those modes $n$ which satisfy one of the following resonance conditions:
\begin{equation}
\begin{aligned}
n &= m, && \text{self-coupling},\\
n &= m \pm N_g j, && \text{forward and backward coupling},\\
n &= N_g j - m, && \text{cross coupling to conjugate modes}.
\end{aligned}
\label{eq:modecoupling}
\end{equation}
$j_\text{Max}$ is chosen sufficiently large to ensure convergence. In the present computations, we use $j_\text{Max} \ge 1200$. After truncation to $N_p=1200$ modes, the system reduces to a finite-dimensional eigenvalue problem. Since the coefficients are generally complex, the Fourier amplitudes $h_n,\gamma_n$ and their complex conjugates  $\bar{h}_n, \, \bar{\gamma}_n$ must be treated as independent variables. We therefore introduce a vector,
\begin{equation}
\mathbf{Z} =
\left[
h_1, h_2, h_3, \ldots, h_{N_p},
\bar{h}_1, \bar{h}_2, \bar{h}_3, \ldots, \bar{h}_{N_p},
\gamma_1, \gamma_2, \gamma_3, \ldots, \gamma_{N_p},
\bar{\gamma}_1, \bar{\gamma}_2, \bar{\gamma}_3, \ldots, \bar{\gamma}_{N_p}
\right]^T
\label{eq60d}
\end{equation}
The projected linearized equations (\ref{eq43}) for the film thickness and surfactant concentration amplitudes take the form:
\begin{equation}
\begin{aligned}
\sum_{n=1}^{N_p} \frac{d h_n}{dt}
&=
\sum_{n=1}^{N_p} A_{hh}(m,n)\, h_n
+
\sum_{n=1}^{N_p} A_{h\bar{h}}(m,n)\, \bar{h}_n
+
\sum_{n=1}^{N_p} A_{h\gamma}(m,n)\, \gamma_n
+
\sum_{n=1}^{N_p} A_{h\bar{\gamma}}(m,n)\, \bar{\gamma}_n,\\
\sum_{n=1}^{N_p} \frac{d \gamma_n}{dt}
&=
\sum_{n=1}^{N_p} A_{\gamma h}(m,n)\, h_n
+
\sum_{n=1}^{N_p} A_{\gamma \bar{h}}(m,n)\, \bar{h}_n
+
\sum_{n=1}^{N_p} A_{\gamma \gamma}(m,n)\, \gamma_n
+
\sum_{n=1}^{N_p} A_{\gamma \bar{\gamma}}(m,n)\, \bar{\gamma}_n,    
\end{aligned}
\label{eq61d}
\end{equation}
for $m=1,2,3,…..N_p$. Taking the complex conjugate of equations (\ref{eq61d}) yields two additional evolution equations for $\bar{h}_n$ and  $\bar{\gamma}_n$. The full linearized system may then be written compactly as,
\begin{equation}
    \frac{d \mathbf{Z}}{dt}= \mathbf{A} \, \mathbf{Z}
    \label{eq62d}
\end{equation}
where $\mathbf{A}$ is a $4N_p \times 4N_p$ block matrix composed of the coupling matrices $A_{hh}, A_{h\bar{h}}, A_{h\gamma}, A_{h\bar{\gamma}} $ and $A_{\gamma h}, A_{\gamma \bar{h}}, A_{\gamma \gamma}, A_{\gamma \bar{\gamma}} $ are given by,
\begin{equation}
\mathbf{A} =
\begin{pmatrix}
A_{hh} & A_{h\bar{h}} & A_{h\gamma} & A_{h\bar{\gamma}} \\
\overline{A_{h\bar{h}}} & \overline{A_{hh}} & \overline{A_{h\bar{\gamma}}} & \overline{A_{h\gamma}} \\
A_{\gamma h} & A_{\gamma \bar{h}} & A_{\gamma \gamma} & A_{\gamma \bar{\gamma}} \\
\overline{A_{\gamma \bar{h}}} & \overline{A_{\gamma h}} & \overline{A_{\gamma \bar{\gamma}}} & \overline{A_{\gamma \gamma}}
\end{pmatrix}
\label{eq63d}
\end{equation}

The diagonal blocks corresponding to the self-coupling condition $n=m$ are

\begin{equation}
\begin{aligned}
A_{hh}(m,m) &= P_{10}+iQ_{10}q_n-R_{10}q_n^2-iS_{10}q_n^3+T_{10}q_n^4,\\
A_{h\gamma}(m,m) &= iU_{10}q_n-V_{10}q_n^2,\\
A_{\gamma h}(m,m) &= P_{20}+iQ_{20}q_n-R_{20}q_n^2-iS_{20}q_n^3+T_{20}q_n^4,\\
A_{\gamma\gamma}(m,m) &= iU_{20}q_n-V_{20}q_n^2.
\end{aligned}
\end{equation}

Off-diagonal blocks arise from mode coupling through shifted harmonics $k_1$, $k_2$, and $k_3$ and are constructed from the corresponding Fourier coefficients as shown below,

For $n=k_1=m-jN_g$, if $1\le k_1\le N_p$,
\begin{equation}
\begin{aligned}
A_{hh}(m,k_1) &= P_{1j}+iQ_{1j}q_n-R_{1j}q_n^2-iS_{1j}q_n^3+T_{1j}q_n^4,\\
A_{h\gamma}(m,k_1) &= iU_{1j}q_n-V_{1j}q_n^2,\\
A_{\gamma h}(m,k_1) &= P_{2j}+iQ_{2j}q_n-R_{2j}q_n^2-iS_{2j}q_n^3+T_{2j}q_n^4,\\
A_{\gamma\gamma}(m,k_1) &= iU_{2j}q_n-V_{2j}q_n^2.
\end{aligned}
\label{eq67}
\end{equation}

For $n=k_2=m+jN_g$, if $1\le k_2\le N_p$,
\begin{equation}
\begin{aligned}
A_{hh}(m,k_2) &= \overline{P}_{2j}+i\overline{Q}_{2j}q_n-\overline{R}_{2j}q_n^2-i\overline{S}_{2j}q_n^3+\overline{T}_{2j}q_n^4,\\
A_{h\gamma}(m,k_2) &= i\overline{U}_{1j}q_n-\overline{V}_{1j}q_n^2,\\
A_{\gamma h}(m,k_2) &= \overline{P}_{2j}+i\overline{Q}_{2j}q_n-\overline{R}_{2j}q_n^2-i\overline{S}_{2j}q_n^3+\overline{T}_{2j}q_n^4,\\
A_{\gamma\gamma}(m,k_2) &= i\overline{U}_{2j}q_n-\overline{V}_{2j}q_n^2.
\end{aligned}
\label{eq69}
\end{equation}

For $n=k_3=jN_g-m$, if $1\le k_3\le N_p$,
\begin{equation}
\begin{aligned}
A_{h\bar h}(m,k_3) &= P_{1j}-iQ_{1j}q_n-R_{1j}q_n^2+iS_{1j}q_n^3+T_{1j}q_n^4,\\
A_{h\bar\gamma}(m,k_3) &= -iU_{1j}q_n-V_{1j}q_n^2,\\
A_{\gamma\bar h}(m,k_3) &= P_{2j}-iQ_{2j}q_n-R_{2j}q_n^2+iS_{2j}q_n^3+T_{2j}q_n^4,\\
A_{\gamma\bar\gamma}(m,k_3) &= -iU_{2j}q_n-V_{2j}q_n^2,
\end{aligned}
\label{eq75}
\end{equation}
We note that the sign change in the odd-derivative terms in equations (\ref{eq75}) relative to equations (\ref{eq67} and \ref{eq69}) reflects the fact that the conjugate modes $\bar{h}_n$ and $\bar{\gamma}_n$ carry wavenumber $-q_n$. Consequently, the $n$\text{th} derivative contributes a factor $(-iq_n)^n$ rather than $(iq_n)^n$.

Finally, the stability problem is obtained by assuming normal modes of the form
\begin{equation}
\mathbf{Z} = \mathbf{Z_0} e^{\sigma t},
\end{equation}
which yields the eigenvalue problem
\begin{equation}
\sigma \mathbf{Z_0} = \mathbf{A} \mathbf{Z_0} .
\end{equation}

Here, the real part of $\sigma$ represents the temporal growth rate. The eigenvalues and corresponding eigenvectors of the matrix $A$ are obtained numerically using the \textit{Eigensystem} command in \textsc{Mathematica}. The discretised system yields $4N_p$  eigenvalues and eigenvectors each containing $4N_p$ components.

Unlike the homogeneous case, where each perturbation mode is associated with a single Fourier mode of the form $e^{iqx}$, the presence of spatial heterogeneity couples multiple harmonic modes within each eigenfunction. Consequently, a given eigenmode cannot be associated uniquely with a single wavenumber. To construct the dispersion relation, each eigenmode is therefore assigned a dominant wavenumber corresponding to the Fourier component with the largest amplitude in the associated eigenvector. Specifically, for the $j$-th eigenvector $(Z_0^j)$ of dimension $4N_p$, we compute the absolute value of each component and identify the index of the largest by, 
\begin{equation}
n^* = \text{Arg}\,\max\limits_{1 \le n \le 4N_p} \left| \mathbf{Z_0}^{(j,n)} \right|,
\end{equation}
where $\mathbf{Z_0}^{(j,n)}$ denotes the $n$th component of the $j$th eigenvector.

Since the vector $\mathbf{Z_0}$ consists of four blocks of size $N_p$ each, the dominant index  $n^*$ is mapped to the corresponding physical wavenumber as
\begin{equation}
q =
\begin{cases}
q_{n^*} &  \text{if} \quad 1 \le n^* \le N_p, \\
q_{n^* - N_p}  & \text{if} \quad N_p < n^* \le 2N_p, \\
q_{n^* - 2N_p} & \text{if} \quad 2N_p < n^* \le 3N_p, \\
q_{n^* - 3N_p} & \text{if} \quad 3N_p < n^* \le 4N_p .
\end{cases}
\end{equation}
\begin{figure}
  \centering
  \includegraphics[width=1\textwidth]{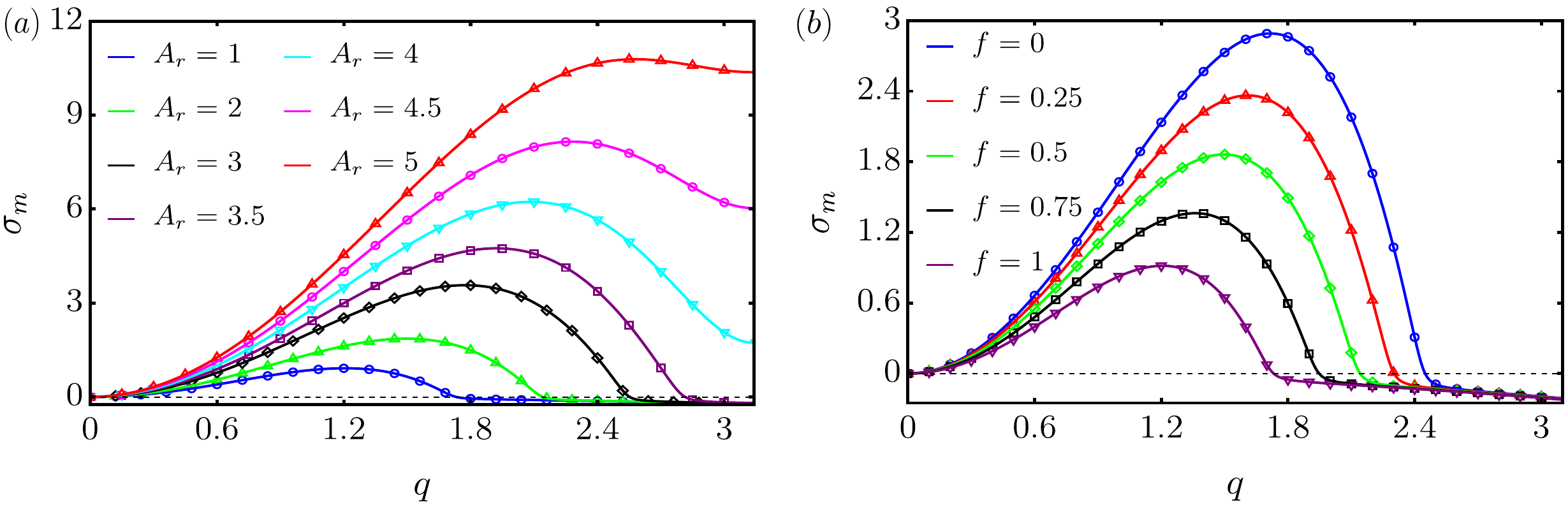}
  \caption{Variation of the perturbation growth rate $\sigma_m$ with wavenumber $q$ for step variation in Hamaker constant. $(a)$ Dispersion curves for different values of $A_r$ at fixed mucin coverage fraction $f=0.5$ $(b)$ Dispersion curves for different mucin coverage fractions $(f)$ at fixed $A_r=2$. The other parameters are $C=1,M=0.1,Pe=100,\beta_0=0.1$, and $\gamma_s=0.5$. The solid lines represent results obtained from Floquet-Bloch theory, while the markers denote predictions from the discretised eigenvalue approach.}
  \label{fig:3}
\end{figure}
The dispersion relation is then constructed by plotting the maximum growth rate $\sigma_m = \max \bigl(\mathrm{Re}(\sigma)\bigr)$
against its dominant wavenumber $q$. Only modes satisfying $\sigma_m \ge -0.25$ are retained in the dispersion plots. Convergence with respect to the truncation parameters $N_p$  and $j_\text{Max}$  is verified by systematically increasing their values until the computed growth rates vary by less than $10^{-5}$.
\subsubsection{Periodic step variation}
 The first Brillouin zone is defined as $q\in [-k_s/2,k_s/2 ]=[-\pi,\pi]$.  This interval represents the physically distinct and non-redundant range of perturbation wavenumbers in the periodic system. In the presence of spatial periodicity, the perturbation eigenmodes satisfy Bloch-wave form given by equation (\ref{eq45}). Bloch modes with quasi-wavenumbers differing by integer multiples of $k_s$ are physically equivalent, since $e^{i (q+\mathbb{Z}k_s)x} \phi(x)=e^{iqx} \phi(x)$. Consequently, modes outside the first Brillouin zone can be mapped back into the interval $q \in [-\pi,\pi]$. Floquet-Bloch theory therefore naturally restricts the independent quasi-wavenumber range to the first Brillouin zone. Modes with $q>\pi$ do not correspond to new physical instabilities. Hence, the dispersion relations are presented only for positive $q\in[0,\pi]$ within the first Brillouin zone. \\
 \begin{figure}
  \centering
  \includegraphics[width=1\textwidth]{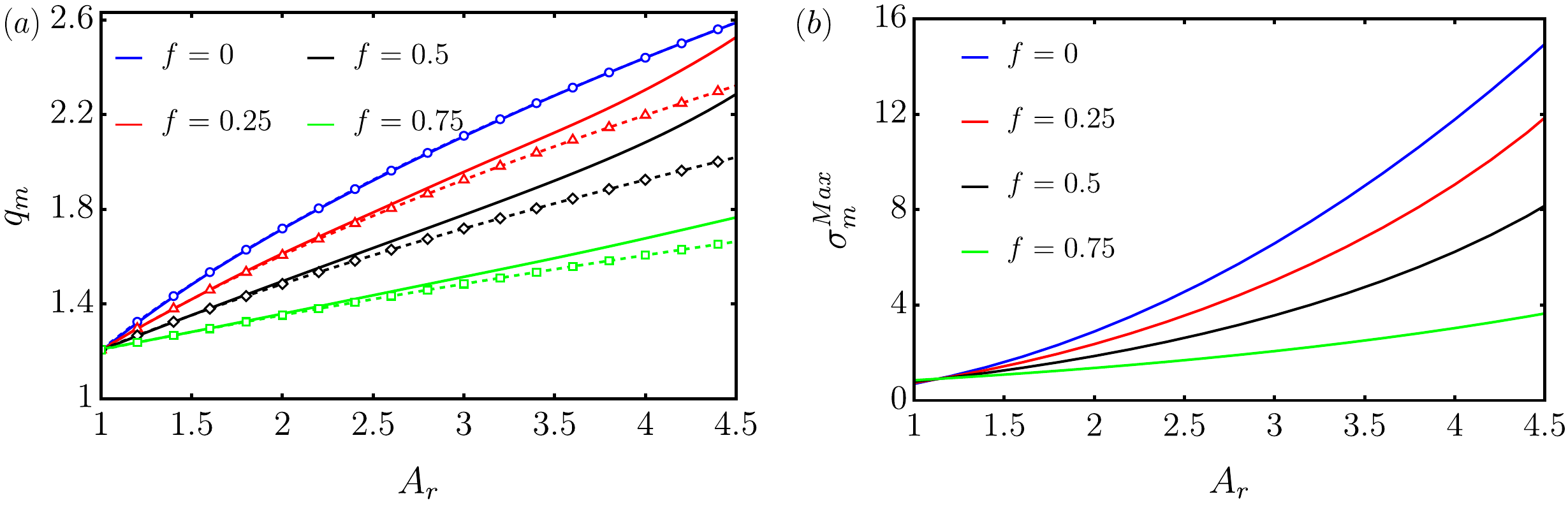}
  \caption{$(a)$ Variation of the most unstable wavenumber $q_m$ with $A_r$ for a periodic step variation. The present results are compared with the approximation of \citep{choudhury2021tear}  $(b)$ Variation of the fastest perturbation growth rate $\sigma_m^{Max}$ with $A_r$ at different mucin coverage fraction $f$. The other parameters are $C=1,M=0.1,Pe=100,\beta_0=0.1,$ and $\gamma_s=0.5.$}
  \label{fig:4}
\end{figure}

 Figure \ref{fig:3}$(a)$ shows the dispersion relation for a periodically varying step change in Hamaker constant with $f=0.5$ for different values of $A_r$. The solid lines represent results obtained from Floquet-Bloch theory, while the markers denote predictions from the discretised eigenvalue approach. Excellent agreement between the two methods is observed over the entire range of wavenumbers. The maximum growth rate increases monotonically with increasing $A_r$. Physically, larger values of $A_r$ correspond to stronger intermolecular attraction in the mucin-deficient regions relative to the mucin-rich regions. This enhances the disjoining pressure gradients within the film. Hence, the destabilizing mechanism responsible for instability increases. As a result, disturbances amplify more rapidly and the tear film becomes more unstable with increasing $A_r$. We observe that the growth rate becomes negative at sufficiently large wavenumbers within the first Brillouin zone $q\in[0,\pi]$ for smaller $A_r$. This indicates the existence of a stable high-wavenumber regime in which short-wavelength disturbances are suppressed by capillary effects. However, the growth rate increases and remains positive throughout the Brillouin zone for $A_r=4,4.5$ and $A_r=5$. This implies that all admissible modes in the first Brillouin zone are unstable for high $A_r$. \\

 The most unstable wavenumber $q_m$ is defined as the value of $q$ at which the growth rate attains its maximum. The cutoff wave number $(q_c )$ is defined as the wave number at which the growth rate changes its sign. Both $q_m$ and $q_c$ depend on the $A_r$ and increase with increasing $A_r$. Physically, this indicates that stronger intermolecular attraction enhances the growth rate of instability. The wave number at which growth rate is maximum also shifts to the right and hence the dominant instability shifts towards shorter wavelengths with characteristic length scale $2\pi/q_m$ . Figure \ref{fig:3}$(b)$ depicts the influence of the mucin coverage fraction $f$ on the stability characteristics for a fixed $A_r=2$. The growth rate of perturbations decreases with increasing $f$ confirming the stabilising effect of mucin coverage. In addition, both $q_m$  and $q_c$  decrease as $f$ increases. Thus, higher mucin coverage not only suppresses instability growth but also shifts the dominant instability towards longer wavelengths.

 Figure \ref{fig:4}$(a)$ shows the dependence of the most unstable wavenumber $q_m$ on $A_r$ for different values of the mucin coverage fraction $f$. The results indicate that $q_m$ increases monotonically with increasing $A_r$. We compare the results using our Floquet theory with those reported by choudhury \textit{et al} \cite{choudhury2021tear} who analysed a similar system. In their study, the effect of heterogeneous $A_k (x)$ was approximated using an effective Hamaker constant $A_e=f+(1-f)A_r$. The stability analysis was then performed using a conventional normal-mode approach by assuming that the mucin is spatially uniform with Hamaker constant $A_e$. In contrast, the present analysis explicitly accounts for spatial heterogeneity. The dashed curves with markers in Figure \ref{fig:4}$(a)$ represent the predictions based on the effective Hamaker constant whereas solid lines denote results from Floquet/discretized eigenvalue method. A good agreement between the two approaches is observed in the limiting case $f=0$. However, significant deviations emerge for intermediate values of $f$, particularly at larger $A_r$. This indicates that the effective-medium approximation becomes progressively less accurate as $A_r$ increases. The largest deviation occurs at $f=0.5$, where the heterogeneity is highest. \\
Classical normal-mode analysis while using $A_e$ has spatially homogeneous coefficients and therefore treats each wavenumber independently. The normal mode analysis neglects the interactions between modes and this is captured by Floquet-Bloch theory or discretised eigenvalue methods when the linearised system has periodic coefficients. These interactions fundamentally modify the stability characteristics of periodically  heterogeneous system. We conclude that effective or spatially averaged descriptions of wettability are insufficient to accurately capture the linear stability characteristics of the tear film.\\
\begin{figure}
  \centering
  \includegraphics[width=1\textwidth]{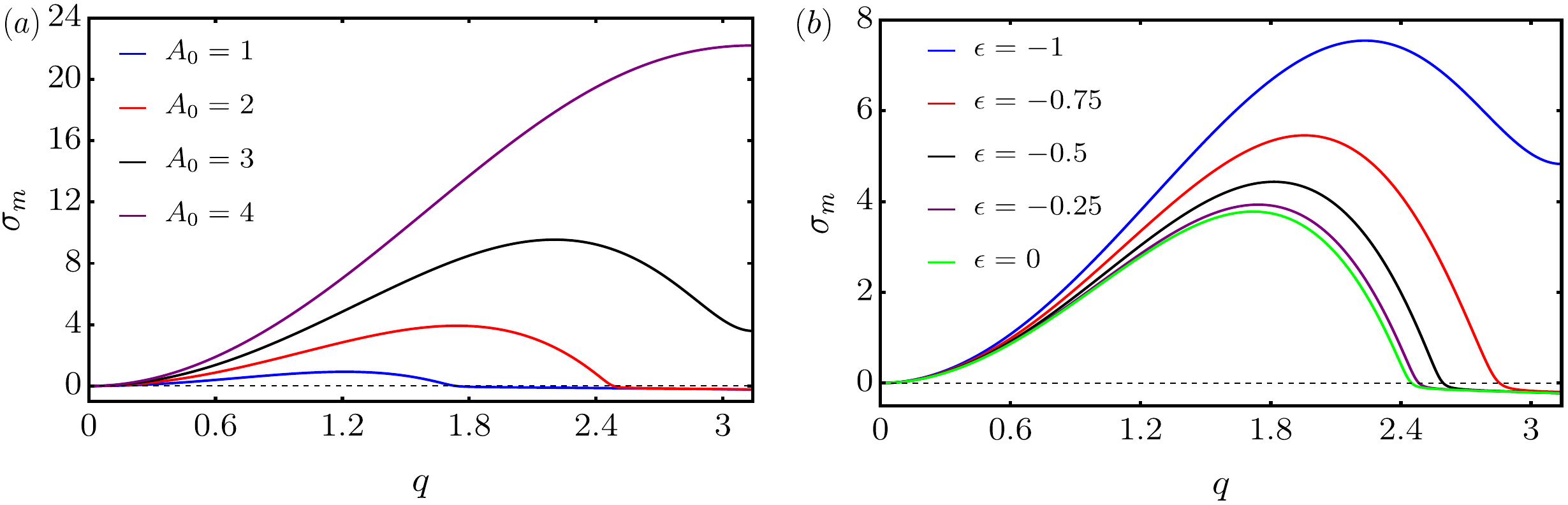}
  \caption{$(a)$ Dispersion curves showing the variation of the perturbation growth rate $\sigma_m$ with wavenumber $q$ for different values of the $A_0$ in the case of a sinusoidally varying $A_k(x)$ with $\epsilon=-0.25$ $(b)$ Dispersion curve for different values of $\epsilon$ at $A_0=2$. The other parameters are $C=1,M=0.1,Pe=100,\beta_0=0.1,$ and $\gamma_s=0.5.$}
  \label{fig:5}
\end{figure}
Figure \ref{fig:4}$(b)$ illustrate the variation of  $\sigma_m^{Max} = \max(\sigma_m)$
 with $A_r$ for different mucin coverage fraction $(f)$. $\sigma_m^{Max}$ decreases monotonically with increasing $f$. Physically, the  extent of  mucin-deficient regions where destabilising van der Waals forces are strongest decreases. As a result, the disjoining-pressure-driven instability is progressively weakened and the tear film becomes more stable. This trend is also consistent with clinical observations that enhanced mucin coverage promotes tear film stability.
\subsubsection{Sinusoidal variation}
We now discuss the stability analysis of a sinusoidally varying Hamaker constant $A_k (x)$. Here, the base state film profile $h_s (x)$ is spatially periodic, as discussed in Section \ref{sec:LSA}. Figure \ref{fig:5}$(a)$ shows the variation of the growth rate with wavenumber $q$ in the first Brillouin zone $q\in[0,\pi]$.
  \begin{figure}
  \centering
  \includegraphics[width=1\textwidth]{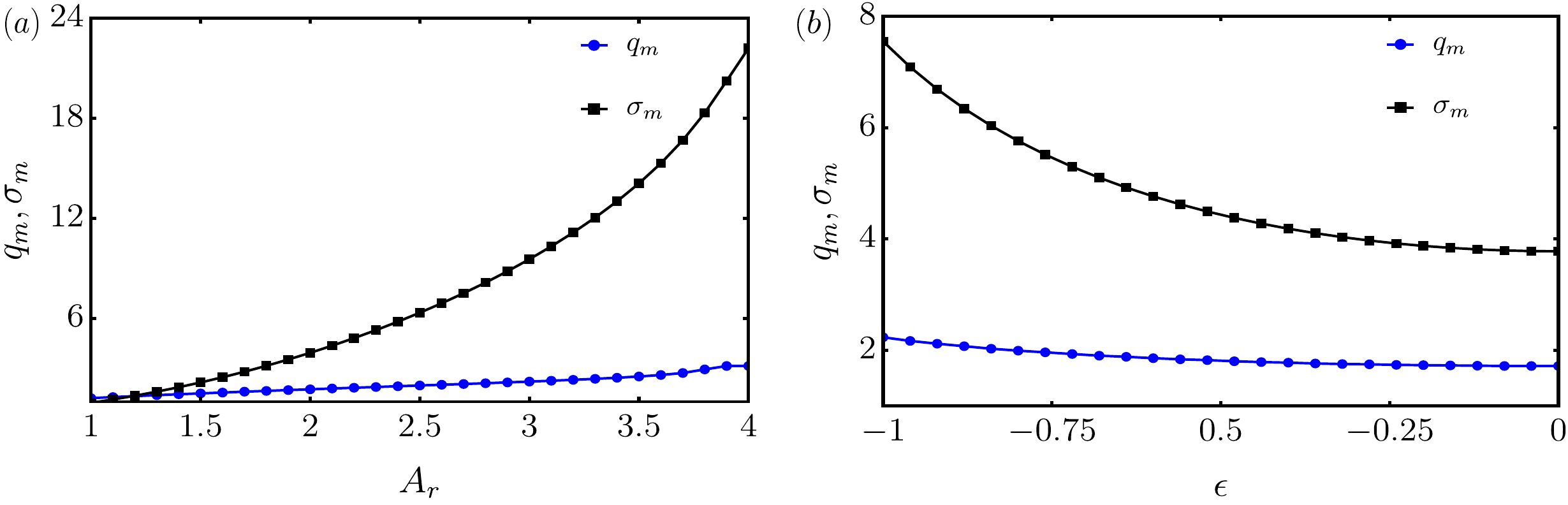}
  \caption{Variation of the most unstable wave number $q_m$ and fastest growth rate $\sigma_m^{Max}$ for a sinusoidally varying Hamaker constant $(a)$ Effect of the  $A_0$ at fixed heterogeneity amplitude  $\epsilon=-0.25$ (b) Effect of the heterogeneity amplitude  $\epsilon$  at $A_0=2$. The other parameters are $C=1,M=0.1,Pe=100,\beta_0=0.1,$ and $\gamma_s$=0.5.  }
  \label{fig:6}
\end{figure}
The growth rate increases with increasing mean amplitude $A_0$. This arises from stronger intermolecular attraction in mucin-deficient regions which has a destabilising influence. Here, both the cutoff wavenumber $q_c$  and the most unstable wavenumber $q_m$  increase with $A_0$. Higher mucin heterogeneity not only enhances the growth rate but it also shifts the dominant instability towards shorter wavelengths. For sufficiently large  $A_0$, the growth rate remains positive throughout the Brillouin zone $q \in [0,\pi]$, implying that all admissible modes become unstable. Figure \ref{fig:5}$(b)$ shows the dispersion curves for different values of the heterogeneity amplitude $\epsilon$. The growth rate of perturbation increases with $|\epsilon|$.  \\
Figure \ref{fig:6}($a$, $b$) shows the variation of the most unstable wavenumber $q_m$ and the maximum growth rate $\sigma_m^{Max}$ with $A_r$ and $\epsilon$ respectively. Both quantities increase monotonically with increasing $A_0$ and $|\epsilon|$. Physically, this indicates that stronger intermolecular attraction together with enhanced spatial heterogeneity in mucin concentrations amplifies the instability. Consequently, perturbations grow more rapidly and the dominant instability shifts towards shorter wavelengths.

\section{Nonlinear numerical simulation}
\label{sec:nonlin}
\subsection{Step variation in Hamaker constant}
\begin{figure}
  \centering
  \includegraphics[width=0.6\textwidth]{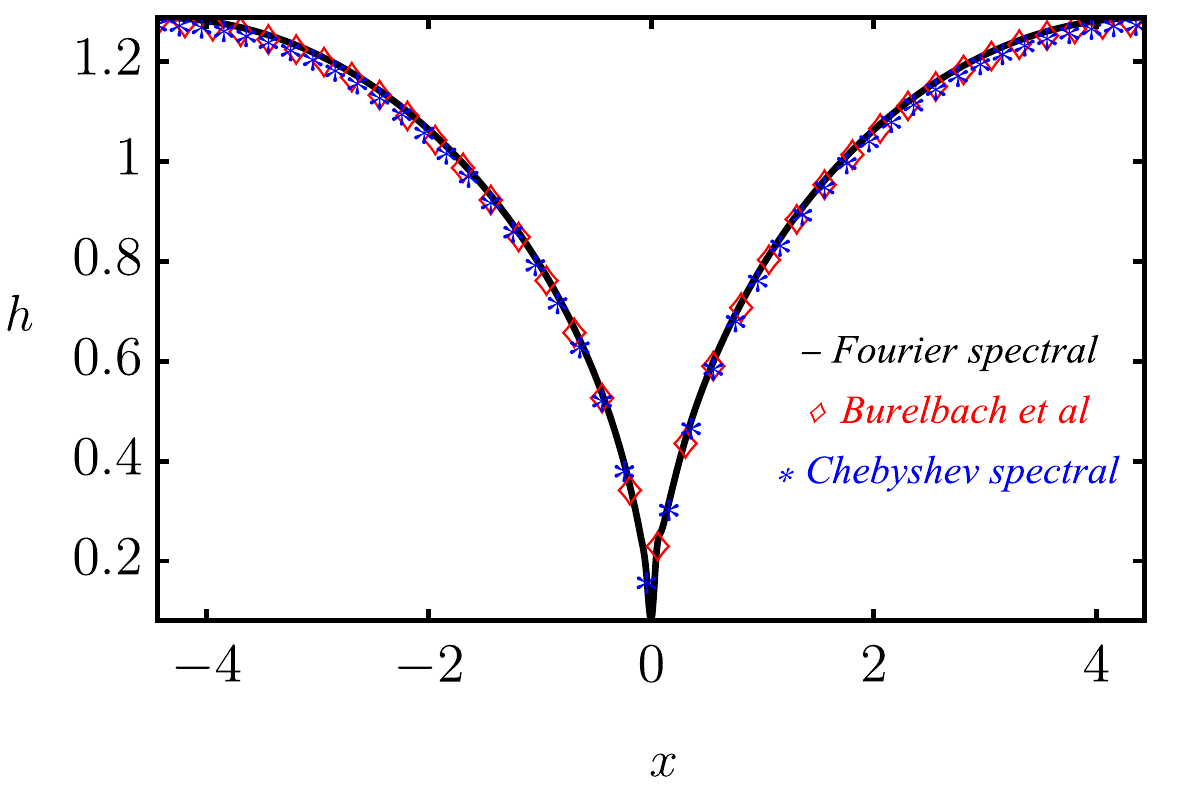}
  \caption{Validation of our in-house numerical scheme using the Fourier spectral method and the Chebyshev spectral method implemented in \textit{Mathematica}. The figure shows the film profile at the time of rupture, demonstrating excellent agreement between the two methods and the result of burelbach \textit{et al} \citep{burelbach1988nonlinear}. }
  \label{fig:7}
\end{figure}

The linear stability analysis identifies the most unstable wavenumber $q_m$ and the corresponding characteristic wavelength $\lambda_m=2\pi/q_m$. However, it does not capture the nonlinear evolution of the tear film. The computational domain is chosen as $[0,\Omega]$ where $\Omega = \lambda_m$. This allows one full wavelength of the fastest-growing disturbance to develop in the doamin. The coupled governing equations (\ref{eq35}-\ref{eq36}) are solved numerically using a Fourier spectral collocation method subject to the periodic boundary conditions, 
\begin{equation}
\begin{aligned}
&h(t,0) = h(t,\Omega), 
\qquad
h'(t,0) = h'(t,\Omega), \\
&h''(t,0) = h''(t,\Omega), 
\quad
h'''(t,0) = h'''(t,\Omega), \\
&\gamma(t,0) = \gamma(t,\Omega), 
\qquad
\gamma'(t,0) = \gamma'(t,\Omega).
\end{aligned}
\label{eq53}
\end{equation}
The spatial discretisation employs $N_p=100$ uniformly distributed grid points,\\
\begin{equation}
x_j = \frac{j \Omega}{N_p}, 
\qquad
j = 0,1,2,\ldots, N_p - 1.
\label{eq52}
\end{equation}
Spatial derivatives are evaluated using Fourier differentiation matrices as described in section \ref{sec:LSA}. The resulting spatial discretisation reduces the governing equations to a system of nonlinear ordinary differential equations in time, which is integrated using the adaptive time-stepping solver \textit{NDSolve} in \textsc{Mathematica}.\\
The numerical framework developed is validated with the isothermal thin-film rupture problem in the absence of surfactant for homogeneous mucin coverage, analysed  by  burelbach \textit{et al} \citep{burelbach1988nonlinear}. Using the Fourier and Chebyshev  spectral methods, the predicted rupture time is $t_{\text{rup}}=4.085$ and $t_{\text{rup}}=4.078$ respectively. Both values are in close agreement with the reported result $t_{\text{rup}}=4.164$ reported in the literature. The film profiles at rupture in Figure \ref{fig:7}, also  show excellent agreement with the reported solution. All subsequent simulations are performed using the Fourier spectral method due to its computational efficiency and high spatial accuracy for periodic domains.  \\

Here, the prime $(')$ denotes differentiation with respect to $x$. The initial condition for the film thickness is taken as a small perturbation about the steady state, $h(t=0)=h_s(x) +H_0 \cos (q_m x)$ where $h_s(x)$ is the steady-state film thickness and $q_m$ is the most unstable wavenumber obtained from the linear stability analysis. The initial surfactant concentration is taken to be uniform, $\gamma(t=0)=\gamma_s$. Unless otherwise stated, the following baseline parameter values are used $M=1,Pe_s=100,\beta_0=0.1,C=1,\gamma_{s}=0.5$,  and $H_0=0.01$. \\
\begin{figure}
  \centering
  \includegraphics[width=1\textwidth]{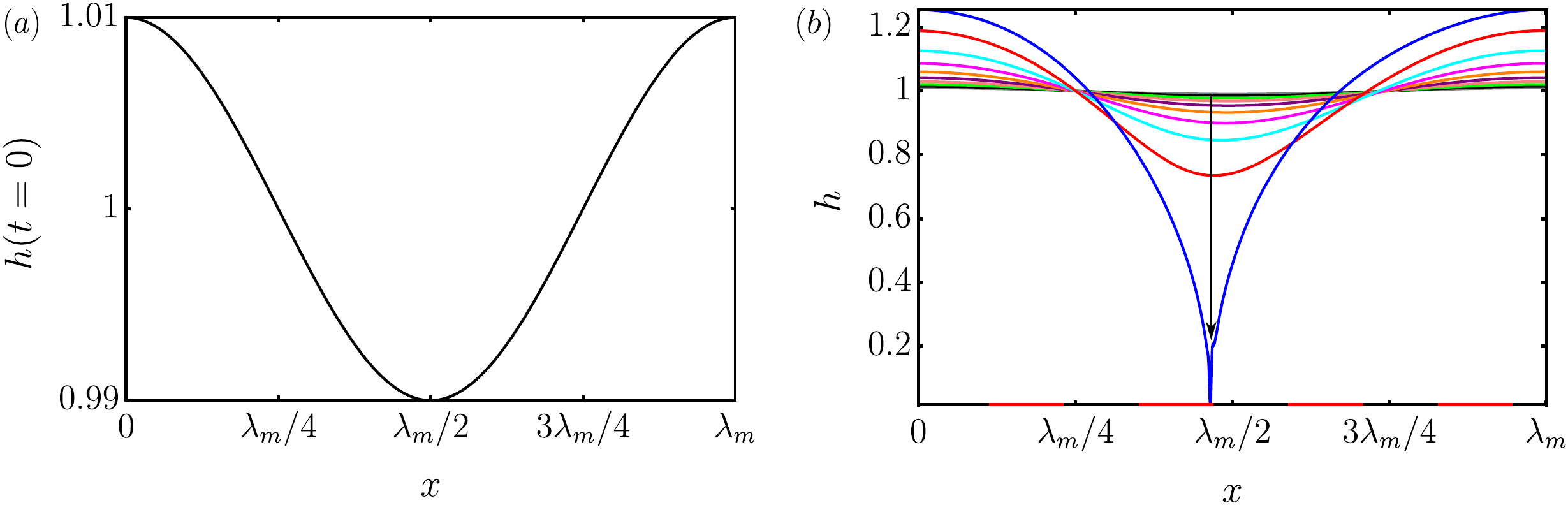}
  \caption{ $(a)$ Initial film thickness over the corneal surface for a step variation in $A_k (x)$ $(b)$ temporal evolution of the tear film thickness obtained from numerical solution of equations (\ref{eq35}-\ref{eq36}). The initial condition is specified as $h =h_s(x)+H_0 \cos(q_m x),\gamma=\gamma_s$. The arrow indicates the progression of time from $t=0$ till the rupture time $t_\text{{rup}}$. The other parameters are $C=1,M=0.1,Pe=100,\beta_0=0.1,f=0.5,A_r=2,h_s(x)=1$ and $\gamma_s=0.5$. The red and black regions on the $x$-axis denote regions of higher and lower $A_k(x)$, respectively. }
  \label{fig:8}
\end{figure}
For the step variation in the Hamaker constant, the steady state film profile is $h_s (x)=1$. For the chosen parameter values, the most unstable wavenumber is $q_m=1.496$ at $A_r=2$. The corresponding growth rate is $\sigma_m=2.1$. This indicates that the system is unstable. Figure 8$(a)$ shows the initial film profile used in the simulations, which attains a minimum at $x=\lambda_m/2$. The temporal evolution of the film thickness is shown in Figure 8$(b)$ at time intervals of $\delta t=0.2$. Although the initial minimum is located at $x=\lambda_m/2$, it progressively shifts towards the left as the film evolves. This shift is induced by the spatial variation in $A_k (x)$. The red and black regions on the $x$-axis in Figure \ref{fig:8}$(b)$ denote regions of higher and lower $A_k (x)$, respectively. The point $x=\lambda_m/2$  lies within a mucin-rich region characterised by lower Hamaker constant $(A_{k1}=1)$, whereas the adjacent mucin-deficient region (shown in red) has larger Hamaker constant $(A_{k2}=2)$. The disjoining pressure is stronger in the mucin-deficient region. Hence, the attraction between the tear film and corneal surface is high in this region. This causes the corresponding thinning location to shift over time. The instability is driven by van der Waals forces, which scale as $\sim h^{-3}$. As the film thins, these forces increase rapidly in magnitude. This amplifies the perturbation further and accelerate the thinning process. Consequently, rupture occurs in the mucin-deficient region at $t_\text{rup}=1.73$ and $x=1.96$ where $A_{k2}=2$. The numerically computed film thickness $h$ is used to obtain the velocity field within the tear film. \\

\begin{figure}
  \centering
  \includegraphics[width=0.7\textwidth]{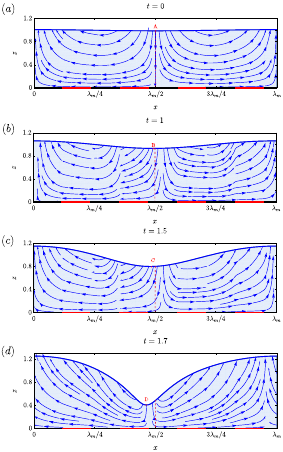}
  \caption{ Streamlines showing the evolution of the velocity field within the tear film at different time instants for a step variation in Hamaker constant $A_k (x)$. The other parameters are $C=1,M=0.1,Pe=100,\beta_0=0.1,A_r=2,f=0.5$ and $\gamma_s=0.5$. }
  \label{fig:9}
\end{figure}
Figure \ref{fig:9} represents the velocity streamlines within the tear film at four different time instants. At $t=0$, the flow is relatively weak and symmetric about $x=\lambda_m/2$. The motion is initially driven by pressure gradients generated by the imposed perturbation. As the film evolves, fluid is progressively redistributed in response the pressure and the disjoining-pressure gradients. Van der Waals forces drive fluid away from the thinner regions (valleys) towards the thicker regions (crests), causing the valleys to thin further with time. This enhances the local pressure gradients and strengthens the flow directed away from the thinning region. As time progresses, the spatial heterogeneity makes the flow increasingly asymmetric. The fluid is preferentially drawn away from the mucin-deficient region in the vicinity of $x=\lambda_m/2$ where the disjoining pressure is stronger due to the larger Hamaker constant. This results in a lateral shift of the thinning region towards the left. Consequently, the location of the minimum film thickness shifts progressively over time (see points A-D in Figure \ref{fig:9}($a$-$d$)) and finally ruptures at $x=1.96$ which lies in the region of low wettability. The inward motion of the liquid is accompanied by the development of counter-rotating vortical structures particularly near $x=\lambda_m$. This arises as the flow must satisfy the imposed periodic boundary conditions. As the film approaches rupture, these vortices become more pronounced as observed at $t=1.7$ in Figure \ref{fig:9}$(d)$. These results confirm that spatial heterogeneity in wettability plays an important role in governing the dynamics of tear film rupture. In particular, the film ruptures within the regions of reduced wettability irrespective of the location of  minimum of the initial disturbance. The local amplification of disjoining pressure within mucin-deficient regions, promotes preferential thinning and accelerates rupture at these locations. This is consistent with clinical observations, where tear film breakup is repeatedly observed at the same spatial locations following successive blinks \citep{tong2021assessment,tsubota2020new}. In contrast, homogeneous wettability models typically predict rupture at the location of the minimum of initial film thickness \citep{zhang2003surfactant} and therefore cannot capture the  localisation of rupture identified in the present study.\\

\begin{figure}
  \centering
  \includegraphics[width=1\textwidth]{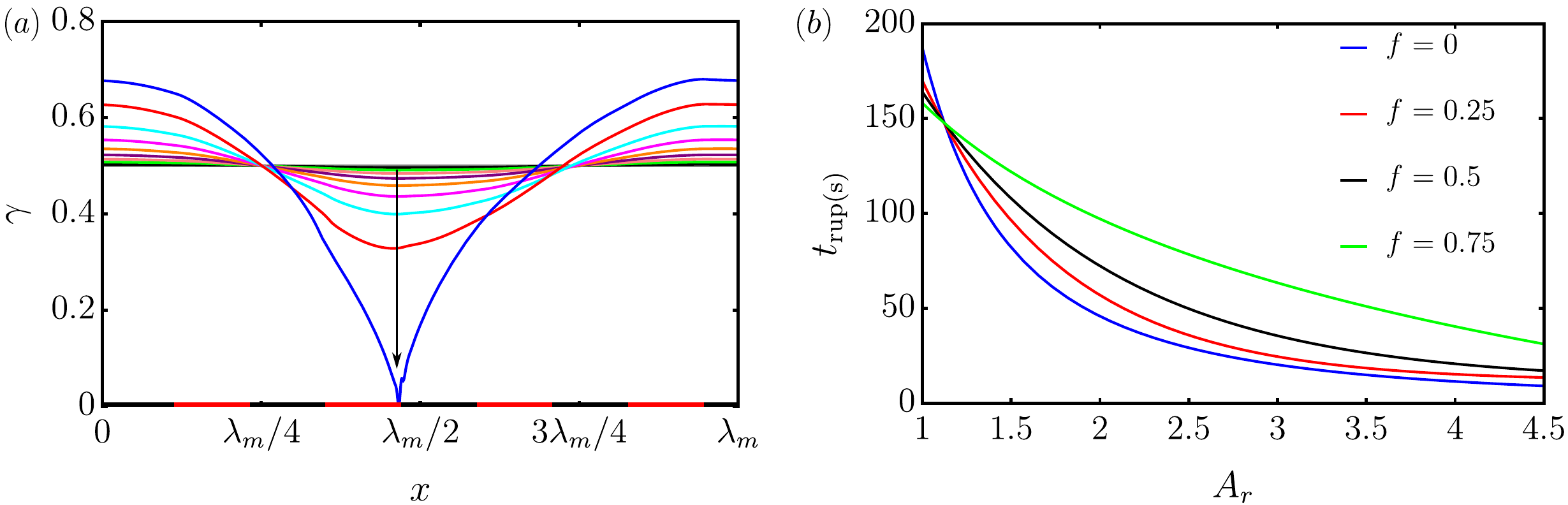}
  \caption{$(a)$ Temporal evolution of the lipid concentration obtained from the nonlinear numerical simulation for $A_r=2$ and $f=0.5$ $(b)$ Dependence of the dimensional tear rupture time $t_\text{{rup}}$ on $A_r$. The other parameters are $C=1,M=0.1,Pe=100,\beta_0=0.1,$ and $\gamma_s=0.5$.}
  \label{fig:10}
\end{figure}

The accompanying fluid flow transports lipid molecules from the thinning regions (valleys) towards the thicker regions (crests). This leads to depletion of lipid concentration in the thinner regions and accumulation in the thicker regions, as shown in Figure \ref{fig:10}$(a)$. The resulting non-uniform surfactant distribution generates surface tension gradients with higher surface tension in the valleys  and lower surface tension in the crests. These gradients give rise to Marangoni stresses that drive fluid from the crests toward the valleys, opposing the flow induced by van der Waals forces. Consequently, lipid redistribution acts as a stabilizing mechanism that resists further thinning of the tear film. In addition, capillary forces arising from curvature gradients also oppose deformation and act to suppress rupture. However, these stabilizing Marangoni and capillary effects are insufficient to counterbalance the dominant van der Waals attraction. As a result, the film continues to thin and ultimately ruptures when the minimum film thickness approaches the corneal surface. Following rupture, exposure of the underlying surface leads to a locally hydrophobic region, which suppresses rewetting and results  in  the formation of dry patches on the cornea. \\

Figure \ref{fig:10}$(b)$ shows the dependence of the tear film rupture time on $A_r$ for different mucin coverage fractions $f$. The initial perturbation in each case is chosen based on the most unstable wavenumber $q_m$, which depends on $A_r$. For a fixed $f$, increasing $A_r$ leads to a reduction in rupture time as stronger intermolecular attraction accelerates film rupture. In contrast, the rupture time increases with increasing mucin coverage fraction $f$. This indicates a stabilizing influence of the mucin-rich regions. These trends are consistent with the linear stability predictions, which show that the growth rate of perturbations increases with $A_r$ and decreases with $f$. Over the range $1\leq A_r\leq 4.5$, the rupture time decreases from approximately 187.5 s  to 9 s , as shown in Figure \ref{fig:10}$(b)$. This underscores the strong sensitivity of tear-film stability to mucin coverage. 
\subsection{Sinusoidal variation}
\begin{figure}
  \centering
  \includegraphics[width=1\textwidth]{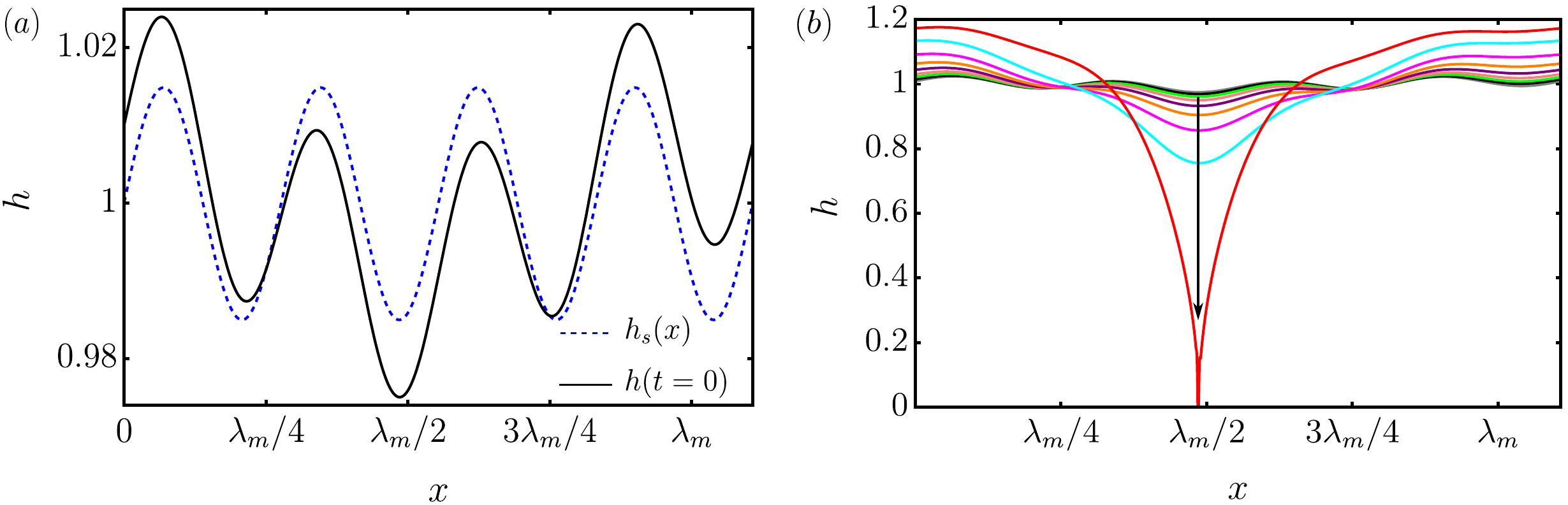}
  \caption{ $(a)$ Steady state film thickness and initial tear film profile over the corneal surface for a sinusoidally varying $A_k (x)$ $(b)$ Temporal evolution of the tear film profile obtained from the nonlinear numerical simulations of equations (\ref{eq35}-\ref{eq36}). The other parameters are $C=1,M=0.1,Pe=100,\beta_0=0.1,A_0=2,\epsilon=-0.25$ and $\gamma_s=0.5$.  }
  \label{fig:11}
\end{figure}

In this case, the steady state is spatially non-uniform. For $A_0 =2$, the most unstable wavenumber is $q_m  = 1.74$, with a corresponding growth rate $\sigma_m  = 3.93$. Two distinct length scales are present in the system: the substrate wavelength arising from the periodic nature of the mucin distribution $\lambda_s=2\pi/k_s=1$ and the dominant instability wavelength $\lambda_m=2\pi/q_m$. To capture the nonlinear evolution, the computational domain $\Omega$ must satisfy $\Omega \geq \lambda_m$ so that at least one full wavelength of the fastest-growing disturbance can develop within the domain. If $\Omega<\lambda_m$, the domain artificially suppresses or distorts the instability leading to incorrect nonlinear evolution. This issue does not arise in the piecewise-constant case because the steady-state solution remains spatially uniform. Accordingly, nonlinear simulations for $h$ and $\gamma$ are performed using a Fourier spectral method on the periodic domain $[0,\Omega]$, where $\Omega$ is chosen as the smallest integer greater than $\lambda_m$. This ensures that the base state remains periodic within the computational domain. It also ensures that the dominant instability mode grows freely in the domain.\\
The steady-state profile $h_s (x)$  is periodic with wavelength $\lambda_s=1$. The imposed disturbance has wavelength $\lambda_m=2\pi/q_m$. Since these two wavelengths are generally incommensurate, the resulting initial condition is not strictly periodic over the computational domain as shown in Figure \ref{fig:11}$(a)$. However, the steady state is periodic in the domain $[0,\Omega]$. Figure \ref{fig:11}$(b)$ shows the temporal evolution of the film thickness at time intervals of $\delta t=0.1$ until the rupture $t_\text{rup}=0.76$ for $A_0=2$. Although, the capillary and van der Waals forces are exactly balanced at the steady state, the imposed perturbation disturbs this balance and generates pressure gradients within the film. Since the steady state is linearly unstable, the disturbance grows with time. This drives progressive thinning of the film and eventually leads to rupture.

 In the present case, the initial minimum in film thickness is located to the left of $x=\lambda_m/2 \approx 1.75$ and rupture is also observed to occur at the same location, as shown in Figure \ref{fig:11}$(b)$. We further note from Figure \ref{fig:2} that the minima of the steady-state profile always occur at a location where $A_k (x)$ is largest. Since the steady-state film profile depends directly on the spatial variation of $A_k (x)$, the location of the initial minimum is determined by the mucin distribution. This demonstrates that the rupture location is strongly governed by the spatial variation in intermolecular forces. This behaviour differs fundamentally from the piecewise-constant case. There, the steady state remains spatially uniform and the thinning region migrates towards the mucin-deficient region during the nonlinear evolution. In contrast, for sinusoidally varying $A_k (x)$, the non-uniform steady state already possesses a minimum at the location of strongest intermolecular attraction. Heterogeneity in $A_k (x)$  therefore preconditions the film, causing rupture to occur directly at the location where initial film thickness is minimum.\\
\begin{figure}
  \centering
  \includegraphics[width=1\textwidth]{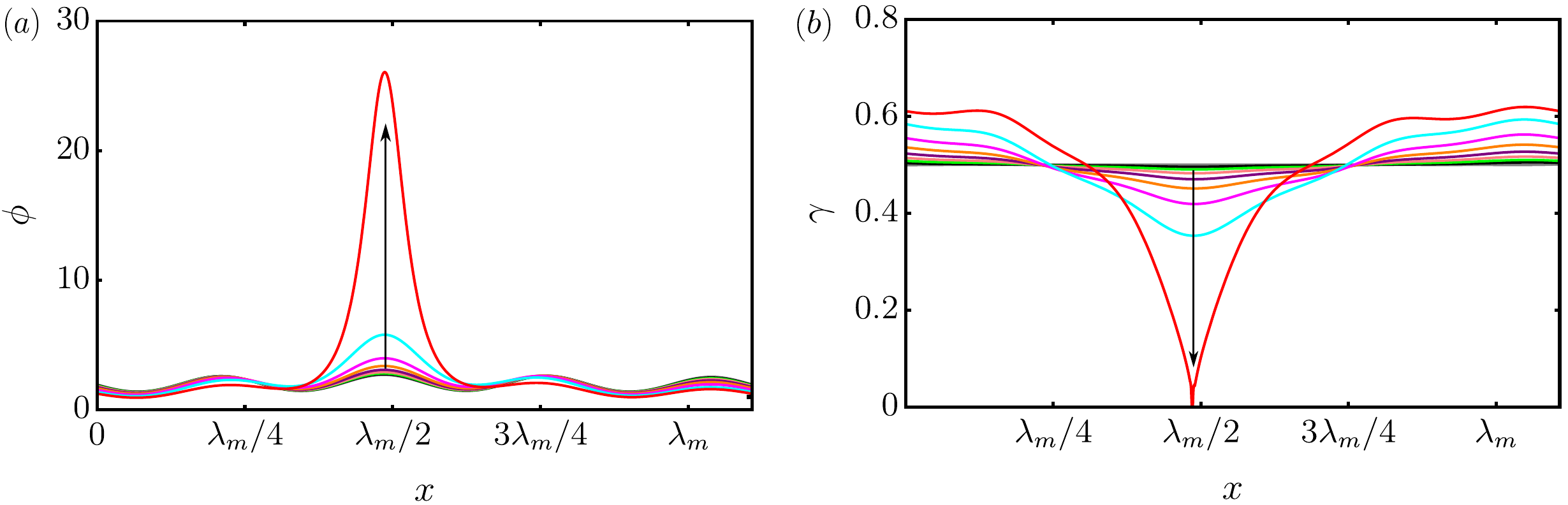}
  \caption{Temporal evolution of  $(a)$ the van der Waals potential $\phi$ $(b)$ the lipid concentration $\gamma$ obtained from the nonlinear numerical simulations for sinusoidal variation in  $A_k(x)$. The other parameters are $C=1,M=0.1,Pe=100,\beta_0=0.1,A_0=2,\epsilon=-0.25$ and $\gamma_s=0.5.$ } 
  \label{fig:12}
\end{figure}
Figure \ref{fig:12}$(a)$ shows the temporal evolution of $\phi$, where a progressive increase is observed near $x \approx 1.75<\lambda_m/2$. Since $\phi$ scales inversely with the cube of the local film thickness $(\sim h^{-3})$, their strength grows rapidly as $h$ decreases over time as shown in Figure \ref{fig:12}$(a)$. Consequently, the rate of thinning accelerates with time. This local amplification of the disjoining pressure enhances thinning in the region near $x\approx 1.75$. This ultimately leads to asymmetric evolution. Figure \ref{fig:12}$(b)$ shows the evolution of the lipid concentration at the tear-air interface. Lipid molecules are advected away from the thinning regions, leading to a depletion in the valleys and accumulation at the crests. \\

\begin{figure}
  \centering
  \includegraphics[width=1\textwidth]{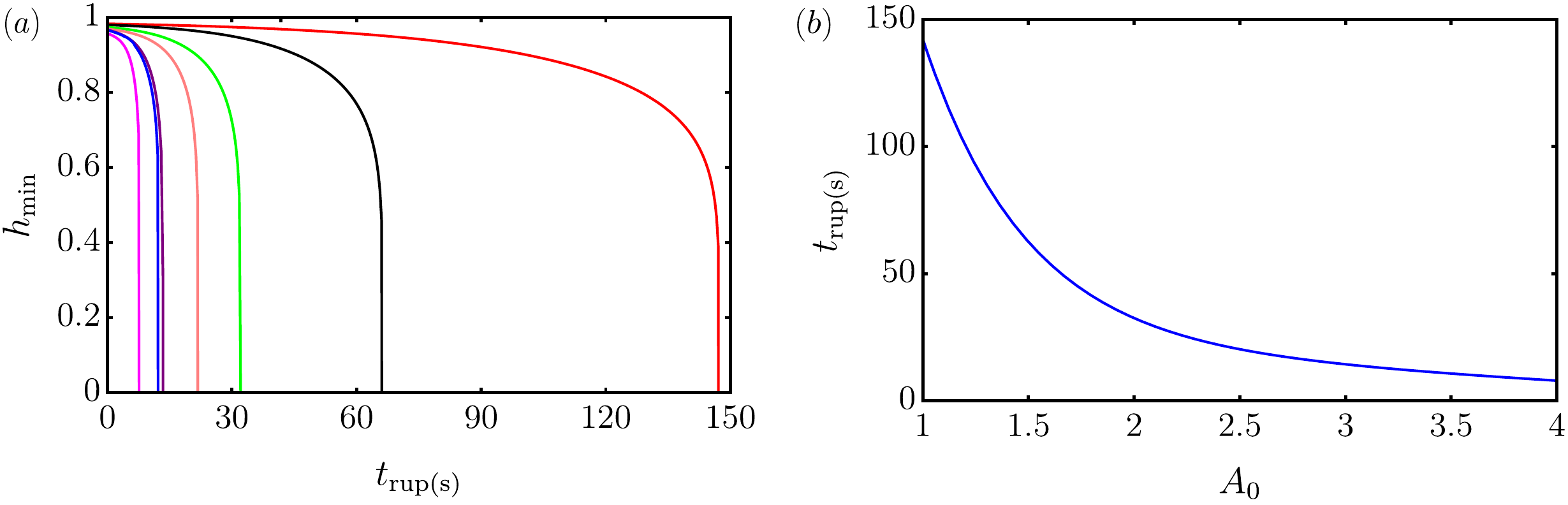}
  \caption{$(a)$ Temporal evolution of the minimum film thickness $h_\text{min}$ for $A_0=1-4$ at the gap of $0.5$ from right to left. $(b)$ Variation of the dimensional rupture time $t_\text{{rup}}$ with $A_r$ for sinusoidal variation in  $A_k(x)$. The other parameters are $C=1,M=0.1,Pe=100,\beta_0=0.1,\epsilon=-0.25$ and $\gamma_s=0.5$.} 
  \label{fig:13}
\end{figure}
Figure \ref{fig:13}$(a)$ shows the temporal evolution of the minimum film thickness $h_{\text{min}}$ for $A_0$=1-4 in increments of $\delta A_0=0.5$ from right to the left for $\epsilon =-0.25$ till rupture. In all cases, the film thickness decreases as the tear film evolves. At early times, the tear film remains relatively thick as the intermolecular attraction between the corneal surface and the tear-air interface is weak. Consequently, capillary forces partially resist thinning. It results in a comparatively slow evolution of the interface. As the film continues to thin, the effective separation between the corneal surface and the tear-air interface decreases. This leads to a rapid amplification of van der Waals attraction.  This creates a positive feedback mechanism in which thinning enhances the attractive force which in turn accelerates further thinning. Consequently, the minimum film thickness decreases very rapidly near rupture as shown in Figure \ref{fig:13}$(a)$. Increasing $A_0$ strengthens the mean intermolecular attraction which accelerates fluid drainage from the thinning regions. As a result, film ruptures significantly faster for larger values of $A_0$ as shown in Figure \ref{fig:13}$(b)$.

Figure \ref{fig:13}$(b)$ shows the corresponding dimensional rupture time $t_{\text{rup}}(\text{s})$ as a function of  $A_0$ for $\epsilon=-0.25$. The rupture time decreases nonlinearly with increasing $A_0$. Larger values of $A_0$ correspond to stronger van der Waals attraction in mucin-deficient regions. This enhances the growth of the instability and accelerates film thinning, leading to significantly shorter rupture times. Physically, these results indicate that even moderate heterogeneity can substantially reduce tear film lifetime. The predicted rupture times fall within physiologically relevant ranges reported experimentally \citep{norn1969desiccation,cho1992tear,korb2001comparison}. In particular, the present model predicts rupture times in the range 8-144 s for $1 \leq A_0 \leq 4$. These are considerably shorter than those typically obtained from homogeneous surfaces. 

\subsection{Comparison with clinical observations}
Clinically, tear breakup time is measured using both invasive and non-invasive techniques. Reported breakup times vary considerably across studies due to difference in environmental conditions, tear composition, blinking patterns, and measurement protocols. Experimental and clinical studies have reported rupture times ranging from as low as 3 s \citep{norn1969desiccation} to  200 s \citep{cho1991stability,korb2001comparison}, with typical values often lying in the range 15-50 s \citep{holly1977tear}. The present heterogeneous wettability model predicts rupture times that fall within these physiologically relevant ranges. For the step variation in wettability, the predicted rupture time decreases from approximately 187.5 s to 9 s as shown in Figure 10(b). Similarly, for sinusoidally varying wettability, the rupture time decreases from approximately 144 s to 8 s, as shown in Figure 13(b). These values match closely with clinically observed breakup times. \\
An additional clinically relevant feature captured by the present model is the localisation of rupture within mucin-deficient regions. This behaviour is qualitatively consistent with clinical observations in which tear film breakup occurs repeatedly at nearly same locations following successive blinks \citep{tong2021assessment,tsubota2020new}. Such behaviour cannot be explained using homogeneous thin-film models, where rupture is generally determined only by the imposed perturbation or the initial minimum film thickness. Overall, the comparison with clinical observations highlights the importance of incorporating spatial heterogeneity in mucin coverage when modelling tear film stability and rupture dynamics.

\section{Conclusions} 
\label{sec:conclusions}
This study has examined the role of spatially heterogeneous mucin coverage in governing the stability,  localisation of rupture and breakup dynamics of the tear film. The investigation is motivated by clinical evidence of non-uniform mucin expression along the corneal surface due to conjunctival goblet cell dysfunction, glycocalyx disruption and increased interfacial friction. To capture these physiological features within a rigorous theoretical framework, a lubrication model has been developed. The spatial periodicity in both the Hamaker constant and the slip length represent mucin-rich and mucin-deficient regions of the corneal epithelium. Linear stability analysis and nonlinear numerical simulation have been employed systematically to explain the influence of both sharply localised and smoothly varying mucin heterogeneity on tear film evolution and rupture. The principal findings of the study are as follows:\\
\begin{enumerate}
  \item 	The nature of the steady state depends strongly on the form of the wettability variation. For step variation in $A_k (x)$, the steady state remains spatially uniform as disjoining pressure gradient vanishes within each region of constant wettability. In contrast, for a smoothly varying $A_k (x)$, the disjoining pressure remains spatially non-uniform even at equilibrium. Consequently, the steady state is determined by a balance between nonzero capillary and van der Waals forces, resulting in a spatially varying base-state film profile. 
  \item 	Linear stability analysis based on Floquet-Bloch Theory and discretized eigenvalue method reveals the mode coupling between different perturbation wavenumbers that is entirely absent in classical normal-mode analyses for homogeneous systems. The analysis demonstrates that the most unstable wavenumber and the corresponding growth rate increase monotonically with increasing wettability difference between the mucin rich and mucin deficient surfaces. The mucin deficiency destabilizes the tear film. 
  \item Nonlinear simulations performed using a Fourier spectral method confirm that rupture consistently occurs within mucin-deficient regions, irrespective of the location of the minimum in initial film thickness. The localization is driven by the local amplification of the disjoining pressure in the regions of reduced mucin coverage. This feature cannot be captured by models assuming spatially uniform surface properties. Furthermore, the rupture time is sensitive to the fraction of region where mucin is present on the corneal surface. 
  \item 	For 4 times increase in the Hamaker constant $(1 \leq A_0 \leq 4)$, the rupture time decreases from approximately 144 s to 8 s, nearly a 18-fold reduction in tear film rupture time.
  \item The present model predicts rupture times that are much shorter than those obtained from homogeneous models and fall within the lower range of clinically observed tear film breakup times.  This arises directly from the localized increase in van der Waals attraction in the mucin-deficient regions.
\end{enumerate}
Overall, the present study demonstrates that heterogeneous wettability plays a decisive role in governing tear film stability, rupture localisation, and rupture times. The theoretical and computational framework developed here provides a systematic methodology for analysing thin film flows over surfaces with spatially periodic, non-uniform wettability, and may find broader application beyond the ocular context. From a physiological standpoint, the present model may be extended incorporating evaporation from the air-tear interface, osmotic transport across the corneal epithelium, and the periodic forcing due to blinking. From an experimental standpoint, direct measurement of the Hamaker constant over mucin-depleted corneal surfaces would enable quantitative comparison with the model predictions reported in this study.
\begin{acknowledgments}
The authors acknowledge the Indian Institute of Technology Madras for providing research facilities. This work was supported by the Prime Minister’s Research Fellowship (PMRF).
\end{acknowledgments}

\appendix

\section{Periodic coefficients in linear stability analysis}
\label{App:A}
The periodic coefficients are given as

\begin{equation}
\begin{aligned}
P_1(x)=&\frac{1}{h_{s}(x)^4}
\Bigg[
h_{s}(x)^2
\Big(
A_k'(x)\big(5h_{s}'(x)-3\beta'(x)\big)
+4A_k(x)h_{s}''(x)
-3\beta(x)A_k''(x)
\Big)
\\[4pt]
&\quad
+2h_{s}(x)
\Big(
h_{s}'(x)
\big(
-4A_k(x)h_{s}'(x)
+6A_k(x)\beta'(x)
+9\beta(x)A_k'(x)
\big)
\\[4pt]
&\qquad\qquad
+6\beta(x)A_k(x)h_{s}''(x)
\Big)
-36\beta(x)A_k(x)\big(h_{s}'(x)\big)^2
\\[4pt]
&\quad
-h_{s}(x)^3A_k''(x)
\Bigg]
\end{aligned}
\end{equation}

\begin{equation}
\begin{aligned}
Q_1(x)=&\frac{1}{h_s(x)^3}
\Bigg[
h_s(x)
\Big(
A_k(x)\big(5h_s'(x)-3\beta'(x)\big)
-6\beta(x)A_k'(x)
\Big)
\\[4pt]
&\quad
+18\beta(x)A_k(x)h_s'(x)
-2h_s(x)^2A_k'(x)
\Bigg]
\end{aligned}
\end{equation}

\begin{equation}
R_1(x)=-\frac{A_k(x)\big(h_s(x)+3\beta(x)\big)}
{h_s(x)^2}
\end{equation}

\begin{equation}
S_1(x) 
=
C\, h_s(x)
\left[
-h_s(x)\big(h_s'(x)+\beta'(x)\big)
-2\beta(x)h_s'(x)
\right]
\end{equation}

\begin{equation}
T_1(x)
=
-\frac{1}{3} C
\left( h_s(x) - \eta f(x) \right)^2
\left( 3\beta - \eta f(x) + h_s(x) \right)
\end{equation}

\begin{equation}
U_1(x)
=
M\, h_s(x)\beta'(x)
+M\big(h_s(x)+\beta(x)\big)h_s'(x)
\end{equation}
\begin{equation}
V_1(x)
=
\frac{1}{2}M\, h_s(x)
\big(h_s(x)+2\beta(x)\big)
\end{equation}
\begin{equation}
\begin{aligned}
P_2(x)
=
&\frac{3\gamma_s}{2h_s(x)^5}
\Bigg[
h_s(x)^2
\Big(
A_k'(x)\big(6h_s'(x)-2\beta'(x)\big)
+4A_k(x)h_s''(x)
\\[4pt]
&\qquad\qquad
-2\beta(x)A_k''(x)
\Big)
+2h_s(x)
\Big(
h_s'(x)
\big(
-6A_k(x)h_s'(x)
\\[4pt]
&\qquad\qquad
+4A_k(x)\beta'(x)
+7\beta(x)A_k'(x)
\big)
+4\beta(x)A_k(x)h_s''(x)
\Big)
\\[4pt]
&\qquad
-32\beta(x)A_k(x)\big(h_s'(x)\big)^2
-h_s(x)^3A_k''(x)
\Bigg]
\end{aligned}
\end{equation}
\begin{equation}
Q_2(x)
=
-\frac{3\gamma_s}{h_s(x)^4}
\Bigg[
h_s(x)
\Big(
A_k(x)\big(\beta'(x)-3h_s'(x)\big)
+2\beta(x)A_k'(x)
\Big)
-7\beta(x)A_k(x)h_s'(x)
+h_s(x)^2A_k'(x)
\Bigg]
\end{equation}
\begin{equation}
R_2(x)
=
-\frac{3\gamma_s A_k(x)\big(h_s(x)+2\beta(x)\big)}
{2h_s(x)^3}
\end{equation}
\begin{equation}
S_2(x)
=
-C\,\gamma_s
\left[
\big(h_s(x)+\beta(x)\big)h_s'(x)
+h_s(x)\beta'(x)
\right]
\end{equation}
\begin{equation}
T_2(x)
=
-\frac{1}{2}C\,\gamma_s\, h_s(x)
\big(h_s(x)+2\beta(x)\big)
\end{equation}
\begin{equation}
U_2(x)
=
M\,\gamma_s
\left(
h_s'(x)+\beta'(x)
\right)
\end{equation}
\begin{equation}
V_2(x)
=
M \gamma_s
\left( \beta  + h_s(x) \right)
+
\frac{1}{Pe}
\end{equation}

\section{Derivation for purely imaginary Floquet exponents}
\label{App:B}
The solution $h_1$ and $\gamma_1$ are required to satisfy periodic boundary conditions on the domain $[0,1]$. We therefore employ a Floquet representation and assume perturbations of the form, $h_1 (x)=e^{\sigma t} e^{\alpha x} \phi(x)$ and $\gamma_1 (x)=e^{\sigma t} e^{\alpha x} \psi(x)$. Here, the Floquet exponent $(\alpha)$ is a complex number. $\phi(x)$  and $\psi(x)$ are periodic with period 1. Imposing periodicity of the perturbations gives $\phi(x+1)=\phi(x)$ and $\psi(x+1)=\psi(x)$. Substituting the Floquet form gives
\begin{equation}
    h_1 (x+1)= e^{\sigma t}e^{\alpha (x+1)}  \phi(x+1)=e^{\sigma t}e^\alpha e^{\alpha x} \phi(x)
\end{equation}
and
\begin{equation}
    \gamma_1 (x+1)=e^{\sigma t}e^{\alpha (x+1)}  \psi(x+1)=e^{\sigma t}e^\alpha e^{\alpha x} \psi(x)
\end{equation}
Hence, periodicity demands $e^\alpha=1$. Therefore, $\alpha=2 \pi in$ where $n \in \mathbb{Z}$ and thus the Floquet exponent $(\alpha)$ is purely imaginary. 
\nocite{*}
\bibliography{apssamp}

\end{document}